%% file: DS3.tex
\documentclass[10pt,journal,compsoc]{IEEEtran}

\ifCLASSOPTIONcompsoc
  \usepackage[nocompress]{cite}
\else
  \usepackage{cite}
\fi

\usepackage{booktabs} 
\usepackage{xcolor}

\usepackage{amsmath}

\usepackage[ruled, linesnumbered]{algorithm2e}
\usepackage{adjustbox}
\usepackage{graphicx}
\usepackage{subcaption}
\usepackage{soul,xcolor}
\usepackage{graphicx}
\usepackage{multirow}
\usepackage{rotating}
\usepackage{array}
\usepackage{enumitem}
\usepackage[normalem]{ulem}
\usepackage{dblfloatfix}
\usepackage[skip=3pt]{caption}
\usepackage{graphics}
\usepackage{ragged2e}
\usepackage[utf8]{inputenc}
\usepackage[T1]{fontenc}
\usepackage{gensymb}

\definecolor{OliveGreen}{rgb}{0,0.6,0}
\definecolor{skyblue}{rgb}{0.06, 0.75, 0.99}

\newcommand{\new}[1]{\textcolor{black}{#1}}

\setstcolor{red}

\begin{document}

\title{\vspace{-2mm}DS3: A System-Level 
	\underline{D}omain-\underline{S}pecific \underline{S}ystem-on-Chip \underline{S}imulation Framework\vspace{-3mm}}

\author{Samet E. Arda, Anish NK, A. Alper Goksoy, Nirmal Kumbhare, Joshua Mack, \\ Anderson L. Sartor, Ali Akoglu, Radu Marculescu and Umit Y. Ogras
\thanks{
\indent S. E. Arda, A. NK, A. A. Goksoy and U. Y. Ogras are with the School of Electrical, Computer and Energy Engineering, Arizona State University, Tempe, AZ 85287 USA. \newline E-mail: \{sarda1, anishnk, aagoksoy, umit\}@asu.edu \newline
\indent N. Kumbhare, J. Mack and A. Akoglu are with the Electrical and Computer Engineering Dept., University of Arizona, Tucson, AZ 85719 USA. \newline E-mail: \{nirmalk, jmack2545, akoglu\}@email.arizona.edu \newline
\indent A. L. Sartor and R. Marculescu are with the Electrical and Computer Engineering Dept., Carnegie Mellon University, Pittsburgh, PA 15213 USA. \newline E-mail: \{asartor, radum\}@cmu.edu}
}
\input{files/0-abstract.tex}

\maketitle

\IEEEdisplaynontitleabstractindextext

\IEEEpeerreviewmaketitle
\input{files/1-introduction.tex}
\input{files/2-related_work.tex}

\input{files/3-overall_goals.tex}
\input{files/4-implementation.tex}

\input{files/5-capabilities.tex}

\input{files/6-validation.tex}
\input{files/7-case_studies.tex}

\input{files/8-conclusion.tex}

\ifCLASSOPTIONcaptionsoff
  \newpage
\fi
\bibliographystyle{IEEEtran}

\vspace{-15pt}
\appendices
\section{\new{DAG Representations of Benchmark Applications}}\label{appendix_DAGs}

\new{Applications from wireless communication and radar domains in the benchmark suite offer a variety of DAG representation ranging from simple one to very complex ones. Figure~\ref{fig:dag_rep} shows DAG representations for WiFi TX/RX, pulse Doppler, and range detection applications. WiFi-TX and WiFi-RX are implemented as five parallel chains of tasks as seen in Figure~\ref{fig:dag_rep}(a) and (b). In addition,
Range detection application only contains seven tasks whereas pulse Doppler is composed of 451 tasks since there are \textit{n} samples for a single signal (see Figures~\ref{fig:radar_bd}(b) and~\ref{fig:dag_rep}(d)).}

\vspace{-10pt}
\ifCLASSOPTIONcompsoc
  \section*{Acknowledgments}
\fi
This material is based on research sponsored by Air Force Research Laboratory (AFRL) and Defense Advanced Research Projects Agency (DARPA) under agreemnet number FA8650-18-2-7860. The U.S. Government is authorized to reproduce and distribute reprints for Governmental purposes notwithstanding any copyright notation thereon. 

The views and conclusion contained herein are those of the authors and should not be interpreted as necessarily representing the official policies or endorsements, either expressed or implied, of Air Force Research Laboratory (AFRL) and Defence Advanced Research Projects Agency (DARPA) or the U.S. Government.

\vspace{-10pt}
\bibliography{references/references}
\vspace{-85pt}
\input{files/biographies.tex}

\end{document}

%% file: files/0-abstract.tex
\IEEEtitleabstractindextext{%
\begin{abstract}
\justifying{
Heterogeneous systems-on-chip (SoCs) are highly favorable computing platforms 
due to their superior performance and energy efficiency potential compared to homogeneous architectures.
They can be further tailored to a specific domain of applications 
by incorporating processing elements (PEs) that accelerate frequently used kernels in these applications.
However, this potential is contingent upon optimizing the SoC for the target domain and utilizing its resources effectively at runtime. 
To this end, system-level design - including scheduling, power-thermal management algorithms and design space exploration studies - plays a crucial role.
This paper presents a system-level domain-specific SoC simulation (DS3) framework to address this need.  
DS3 enables both design space exploration and dynamic resource management for power-performance optimization of domain applications.
We showcase DS3 using six real-world applications from wireless communications and radar processing domain. DS3, as well as the reference applications, is shared as open-source software to stimulate research in this area. 
}
\end{abstract}

\begin{IEEEkeywords}
Heterogeneous computing, SoC, domain-specific SoC, simulation framework, DTPM, design space exploration.
\end{IEEEkeywords}}

%% file: files/1-introduction.tex
\IEEEraisesectionheading{\section{Introduction}\label{sec:introduction}}

%
%
%
%


\IEEEPARstart{H}{omogeneous} general purpose processors provide flexibility to implement a variety of applications and facilitate programmability. 
However, these platforms cannot take advantage of the
domain knowledge to optimize the energy efficiency for specific application domains, such as machine learning, communication protocols, and autonomous driving~\cite{chen2014diannao, reiche2017generating,buzdar2017area}.
In contrast, heterogeneous systems-on-chip (SoCs) that combine general purpose and specialized processors (e.g., audio/video codecs and communication modems) offer great potential to achieve higher efficiency~\cite{hennessy2019new}. 
In particular, domain-specific SoCs (DSSoCs) - a class of heterogeneous architectures - optimize the architecture, computing resources and design flows by exploiting the characteristics of applications in a particular domain. 
For a given target domain, DSSoCs can provide three orders of magnitude higher energy-efficiency in comparison to general-purpose processors~\cite{cong2014architecture}.


Harvesting the potential of DSSoCs depends critically on the integration of optimal combination of computing resources and their effective utilization and management at runtime.
Hence, the first step in the design flow includes analysis of the domain applications to identify the commonly used kernels~\cite{uhrie2019DSSoC}.
This analysis aids in determining the set of specialized hardware accelerators for the target applications.
For example, DSSoCs targeting wireless communication applications obtain better performance with the inclusion of Fast-Fourier Transform (FFT) accelerators.
Similarly, SoCs optimized for autonomous driving applications integrate deep neural network (DNN) accelerators~\cite{liu2017computer}.
Then, a wide range of design- and run-time algorithms are employed to schedule the applications to the processing elements (PEs) in the DSSoC~\cite{chou2008energy, KPN_map_MPSoCs, smit2005run, de2010dynamic}.
Finally, dynamic power and thermal management (DTPM) techniques optimize the SoC for energy efficient operations at runtime. 
Throughout this process, evaluation frameworks, ranging from analytical models and hardware emulation, are needed to explore the design space and ensure that the DSSoC achieves performance, power and energy targets~\cite{cong2014architecture}. 

\begin{figure}[t]
	\centering
 	\includegraphics[width=1.0\linewidth]{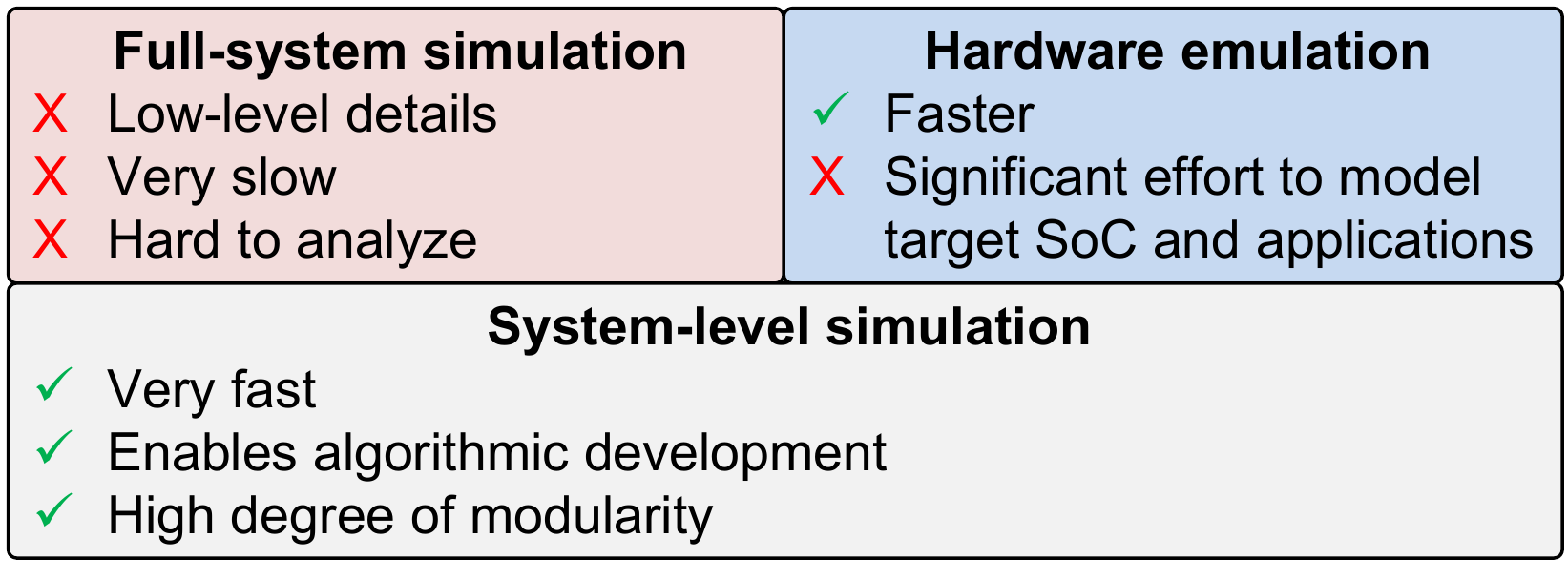}
	\caption{Common DSE methodologies.}
	\label{fig:DSE-methods}
    \vspace{-20pt}
\end{figure}

Full-system simulators, like gem5 \cite{binkert2011gem5}, have the ability to perform instruction-level cycle-accurate simulation. However, this level of detail leads to long execution times, in the order of hours to simulate a few milliseconds of workloads~\cite{sandberg2015full}. 
Hence, they are not suitable for rapid design space exploration. 
It is also important to note that the level of detail provided by cycle-accurate simulations is beyond the requirements of high-level design space exploration.
The most critical system-level questions are 
\textit{where tasks should run} and 
\textit{how fast PEs should operate} to satisfy the design requirements, e.g., maximizing performance per Watt (PPW) or energy-delay product (EDP). 
Contrary to simulation-based approaches, hardware emulation using Field-Programmable Gate Array (FPGA) prototypes are substantially faster~\cite{pellauer2011hasim}. 
However, they involve significantly higher development effort to implement the target SoC and applications. 

Given the design complexities and the cost of considering a large design space, there is a strong need for a simulation environment which allows 
rapid, high-level, simultaneous exploration of scheduling algorithms and power-thermal management techniques, both of which can significantly influence energy efficiency.


In this paper, we present DS3, a system-level domain-specific system-on-chip simulation framework. 
DS3 framework enables 
(1) run-time scheduling algorithm development, 
(2) DTPM policy design, 
and (3) rapid design space exploration. 
To this end, DS3 facilitates plug and play simulation of scheduling algorithms; it also incorporates built-in heuristic and table-based schedulers to aid developers and provide a baseline for users.
DS3 also includes power dissipation and thermal models that enable users to design and evaluate new DTPM policies. 
Furthermore, it features built-in dynamic voltage and frequency scaling (DVFS) governors, which are deployed on commercial SoCs. 
Besides providing representative baselines, 
this capability enables users to perform extensive studies 
to characterize a variety of metrics, PPW and EDP for a given SoC and set of applications. 
Finally, DS3 comes with \textit{six reference applications} from wireless communications and radar processing domain. 
These applications are profiled on heterogeneous SoC platforms, such as Xilinx ZCU-102~\cite{FPGA} and Odroid-XU3~\cite{ODROID}, and included as a benchmark suite in DS3 distribution. 
The benchmark suite enables realistic design space explorations, 
as we demonstrate in this paper.

\vspace{1mm}
\noindent\textbf{The major contributions of this work include}:
\begin{itemize}
    \item A unified, high-level DSSoC simulator, called DS3 that enables design space exploration of hardware configurations together with scheduling and DTPM strategies,

    \item A benchmark suite of real-world applications and their reference hardware implementations and
    
    \item Extensive design space exploration studies for fine-grained architecture tuning.
\end{itemize}

The rest of this paper is organized as follows. The related work is reviewed in Section~\ref{sec:related work}. The goals and architectural details of DS3 are presented in Section~\ref{sec:goals}. Section~\ref{sec:implementation} elaborates the implementation details of DS3 and Section~\ref{sec:capabilities} presents the built-in capabilities. Section~\ref{sec:validation} describes the validation of the framework against a real platform while Section~\ref{sec:case studies} presents the case studies using real-world applications. Finally, the conclusions and future work are discussed in Section~\ref{sec:concandfuture}.

%% file: files/2-related_work.tex
\vspace{-15pt}
\section{Related Work} \label{sec:related work}

\new{
As the use cases for this environment intersect with a large number of distinct research areas, we break the related work into three parts.
First, we discuss existing work in the area of scheduling, power, and thermal optimization algorithms, and we motivate a need for a unified framework that integrates these with rich design space exploration capabilities.
Second, we discuss existing work in the area of design space exploration for embedded systems, and we note the lack of rich support for thermal/power models or plug and play scheduling frameworks.
Third, for completeness, we give a brief overview of related works in the scope of high performance computing or non-embedded environments.
Together, this set of related works serves to motivate the need for an environment such as DS3 that unifies all of these aspects into a single, open-source framework for embedded DSSoC development.
}

\new{Starting with works on scheduling, power, and thermal optimization algorithms,}
one of the most important goals of heterogeneous SoC design 
is to optimize energy-efficiency while satisfying the performance constraints. 
To this end, a variety 
of offline and runtime algorithms have been proposed to schedule applications to PEs in multi-core architectures~\cite{KPN_map_MPSoCs, smit2005run, de2010dynamic,chou2008energy}. 
Similarly, DVFS policies, such as HiCAP~\cite{park2016hicap}, power management governors, such as \textit{ondemand}~\cite{LinuxGovernors}, and thermal management techniques~\cite{coskun2008temperature} have been proposed to efficiently manage the power and temperature of SoCs.
However, existing approaches are typically evaluated in isolated environments and different in-house tools. Hence, there is a strong need for a unified simulation framework~\cite{maurya2018benchmarking}
to compare and evaluate various scheduling and optimization algorithms in a common environment.

\new{Next, there are a large number of works on design space exploration for embedded systems, but they are found to be lacking in support for rich scheduling, thermal, and power optimization algorithms.}
\new{Khalilzad et al.}~\cite{khalilzad2016modularDSE} consider a heterogeneous multiprocessor platform along with applications modeled as synchronous dataflow graphs and periodic tasks. 
The design space exploration problem is solved using a constraint programming solver for different objectives such as deadline, throughput, and energy consumption. ASpmT \cite{neubauer2018exactDSE} proposes a multi-objective tool using Answer Set Programming (ASP) for heterogeneous platforms with a grid-like network template and applications specified as DAGs.
\new{Tr\v{c}ka et al.} \cite{trvcka2011integratedDSE} utilize the Y-chart \cite{kienhuis1997Ychart} philosophy for design space exploration and introduces an integrated framework using the Octopus toolset \cite{Octopus2010} as its kernel module. Then, for different steps in the exploration process (i.e., modeling, analysis, search, and diagnostics), different languages and tools such as Ptolemy, Uppaal, and OPT4J  are employed. Target platforms and applications are modeled in the form of an intermediate representation to support translation from different languages and to different analysis tools. Artemis~\cite{pimentel2001Artemis} aims to evaluate embedded-systems architecture instantiations at multiple abstraction levels. Later, authors extend the work and introduce the Sesame framework \cite{pimentel2006Sesame} in which
target multimedia applications are modeled as Kahn Process Network (KPN) written in C/C++. Architecture models, on the other hand, include components such as processor, buffers, and buses and are implemented in SystemC. The framework supports different schedulers such as first in, first-out (FIFO), round-robin, or customized. A trace-driven simulation is applied for cosimulation of application and architecture models. 


Finally, ReSP~\cite{beltrame2009resp} is a virtual platform targeting multiprocessor SoCs focusing on a component-based design methodology utilizing SystemC and transaction-level modeling libraries. ReSP adopts  lower-level instruction set based simulation approach and is restricted to applications that are already implemented in SystemC. All aforementioned frameworks or tools lack accurate power and thermal models, and do not support for exploration of scheduling algorithms and/or power-thermal management techniques.

\new{Outside of embedded systems, there has also been a large body of work on design space exploration via heterogeneous runtimes at the desktop or HPC scale, with StarPU \cite{augonnet2011starpu} being one of the most prominent examples of such a runtime.}
StarPU is a comprehensive framework that provides the ability to perform run-time scheduling and execution management for directed acyclic graph (DAG) based programs on heterogeneous architectures. Although, the framework allows users to develop new scheduling algorithms, StarPU lacks power-thermal models and DVFS techniques to optimize power and energy consumption. A recent work~\cite{xiao2019self-optimizing} targets domain-specific programmability of heterogeneous architectures through intelligent compile-time and run-time mapping of tasks across CPUs, GPUs, and hardware accelerators. 
In the proposed approach, the authors employ four different simulators, more specifically, 
Contech to generate traces, MacSim to model CPU/GPU architectures, BookSim2 to model the networks-on-chip, and McPat to predict energy consumption. 
The proposed DS3 simulator integrates the above features in a unified framework to benefit similar studies in the future.

\new{To the best of our knowledge, DS3 is the first open-source framework to integrate all of these distinct elements into a unified simulation environment targeting embedded DSSoCs.}
It includes built-in analytical models, scheduling algorithms, DTPM policies, and six reference applications from wireless communication and radar processing domain.


%% file: files/3-overall_goals.tex
\vspace{-15pt}
\section{Overall Goals and Architecture} \label{sec:goals}

The goal of DS3 is to enable rapid development of scheduling algorithms and DTPM policies, while enabling extensive design space exploration. 
To achieve these goals, it provides:

\begin{figure}[b]
    \vspace{-15pt}
	\centering
 	\includegraphics[width=1.0\linewidth]{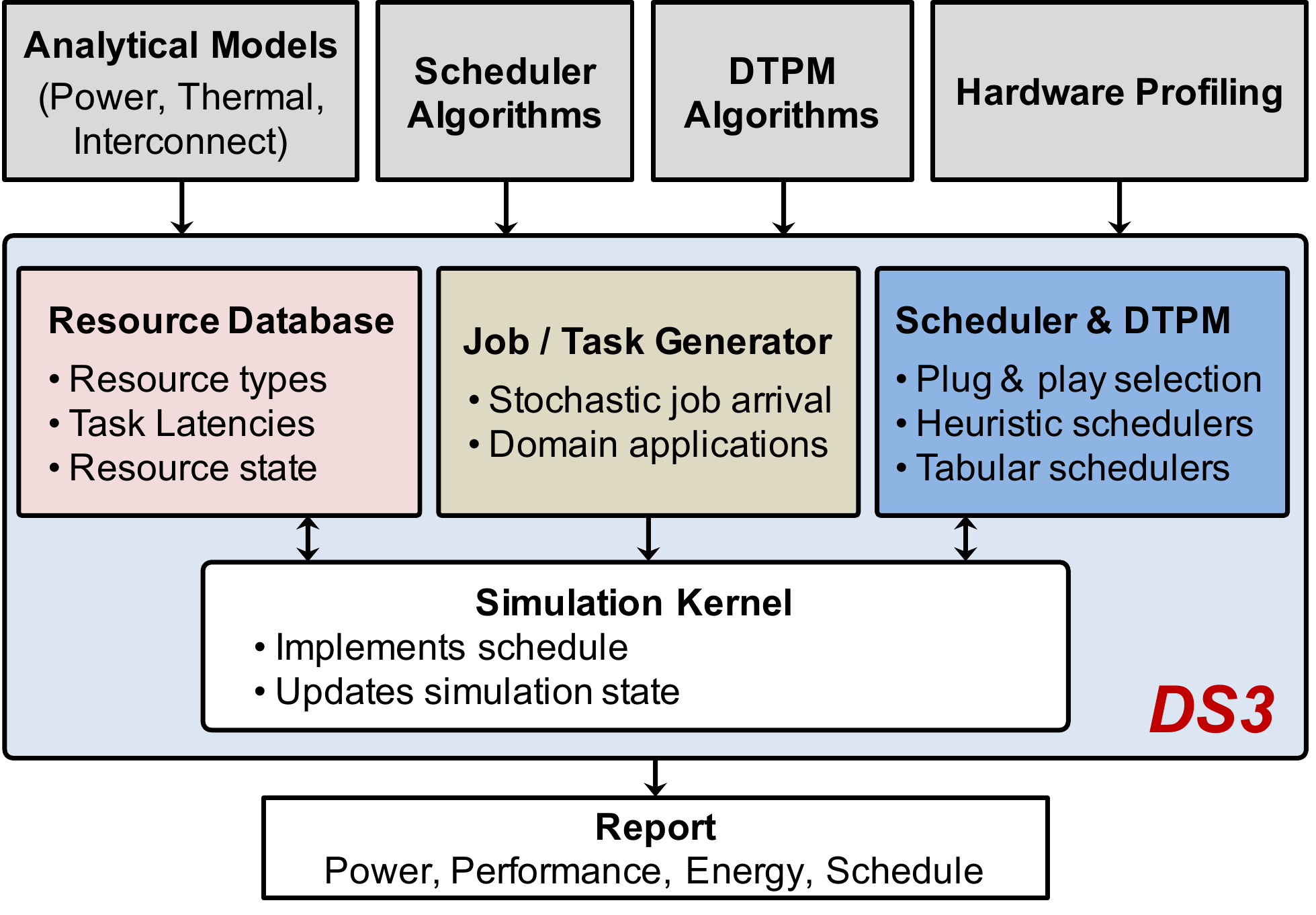}
	\caption{Organization of DS3 framework describing the inputs and key functional components to perform rapid design space exploration and validation.}
	\label{fig:DS3-framework}
	\vspace{-10pt}
\end{figure}

\begin{itemize} [leftmargin=*]
    \item {\bf Scalability:} Provide the ability to simulate instances of multiple applications simultaneously by streaming multiple jobs from a pool of active domain applications.

    \item {\bf Flexibility:} Enable the end-users to specify the SoC configuration, target applications, and the resource database swiftly (e.g., in minutes) using simple interfaces.

    \item {\bf Modularity:} Enable algorithm developers to modify the existing scheduling and DTPM algorithms, and add new algorithms with minimal effort.

    \item {\bf User-friendly Productivity Tools:} Provide built-in capabilities to collect, report and plot key statistics, including power dissipation, execution time, throughput, energy consumption, and temperature.
    
\end{itemize}

The organization of the DS3 framework designed to accomplish these objectives is shown in Figure~\ref{fig:DS3-framework}. 
\new{The resource database contains the list of PEs, including the type of each PE, capacity, operating performance points, among other configurations.}
\new{By exploiting the deterministic nature of domain applications, the profiled latencies of the tasks are also included in the resource database.}
The simulation is initiated by the job generator, 
which generates application representative task graphs.
The injection of applications in the framework is controlled by a random exponential distribution.
The DS3 framework invokes the scheduler at every scheduling decision epoch with the list of tasks ready for execution. 
Then, the simulation kernel simulates task execution on the corresponding PE using execution time profiles based on reference hardware implementations. 
Similarly, DS3 employs analytical latency models to estimate interconnect delays on the SoC~\cite{mandal2019analytical}. 
After each scheduling decision, the simulation kernel updates the state of the simulation, which is used in subsequent decision epochs.
In parallel, DS3 estimates power, temperature and energy of each schedule using power models~\cite{bhat2018algorithmic}.
The framework aids the design space exploration of dynamic power and thermal management techniques by utilizing these power models and commercially used DVFS policies.
DS3 also provides plots and reports of schedule, performance, throughput and energy consumption to help analyze the performance of various algorithms.

DS3 is released to public as a companion to this paper~\footnote{https://github.com/segemena/DS3}. 
Following two sections present the implementation details and capabilities for new developers and users, respectively.

%% file: files/4-implementation.tex
\vspace{-25pt}
\section{Developer View: DS3 Implementation} \label{sec:implementation}

This section describes the implementations of the components of 
DS3 depicted in Figure~\ref{fig:DS3-framework} from a developer's perspective.

\vspace{-15pt}
\subsection{Resource Database} \label{sec:resource_database}

DS3 enables instantiating a wide range of SoC configurations with different types of general- and special-purpose PEs. 
A list of PEs and its characteristics are stored in the resource database.
Each PE in the database has the \textit{static} and \textit{dynamic attributes} described in Table~\ref{tab:pe_attributes}.

\begin{table}[h]
\centering
\caption{List of PE attributes in resource database}
\label{tab:pe_attributes}
\renewcommand{\arraystretch}{1.15}
\begin{tabular}{lcl}
\toprule
 & \textbf{Attribute} & \textbf{Description} \\ \hline
\multirow{6}{*}{\textbf{\rotatebox[origin=c]{90}{Static}}} & \textbf{Type} & \begin{tabular}[c]{@{}l@{}}Defines type of PE\\ Example: CPU, accelerator etc.\end{tabular} \\ \cline{2-3} 
 & \textbf{Capacity} & \begin{tabular}[c]{@{}l@{}}Number of simultaneous threads\\ a PE can execute\end{tabular} \\ \cline{2-3} 
 & \textbf{DVFS policy} & \begin{tabular}[c]{@{}l@{}}Policy which controls PE\\ frequency and voltage at runtime\end{tabular} \\ \cline{2-3} 
 & \textbf{\begin{tabular}[c]{@{}c@{}}Operating \\ performance point\end{tabular}} & \begin{tabular}[c]{@{}l@{}}Operating frequencies and\\ corresponding voltages for a PE\end{tabular} \\ \cline{2-3} 
 & \textbf{\begin{tabular}[c]{@{}c@{}}Execution time\\ profile\end{tabular}} & \begin{tabular}[c]{@{}l@{}}Defines the execution time of\\ supported tasks on each PE\end{tabular} \\ \cline{2-3} 
 & \textbf{\begin{tabular}[c]{@{}c@{}}Power consumption\\ profile\end{tabular}} & \begin{tabular}[c]{@{}l@{}}Provides the power consumption\\ profile of each PE\end{tabular} \\ \hline
\multirow{3}{*}{\textbf{\rotatebox[origin=c]{90}{Dynamic}}} & \textbf{Utilization} & \begin{tabular}[c]{@{}l@{}}Defines active time of a PE for\\ a particular time window\end{tabular} \\ \cline{2-3} 
 & \new{\textbf{Blocking}} & \begin{tabular}[c]{@{}l@{}}\new{Probability that a PE is busy} \\ \new{when a task is ready}\end{tabular} \\ \cline{2-3} 
 & \textbf{State} & \begin{tabular}[c]{@{}l@{}}Indicates whether a PE\\ is busy or idle\end{tabular} \\ \bottomrule
\end{tabular}
\end{table}


\new{The \textit{static} and \textit{dynamic} attributes are determined based on the current industry practice and a careful examination of available systems on the market. For example, CFS scheduler, the default linux kernel scheduler~\cite{CFS}, makes task mapping decisions based on the utilization of PEs. In addition, ARM big.LITTLE architecture~\cite{bhat2018algorithmic}, combining Cortex-A15 cluster with energy-efficient Cortex-A7 cluster, supports different operating frequencies for each cluster. The voltage level and thus energy consumption depends on the operating point and DS3 takes these effects into account. Finally, commercial SoCs utilize DVFS policies~\cite{LinuxGovernors} to control power and performance of PEs. For this reason, we integrated these policies into DS3 and assigned the current DVFS policy as an attribute to a PE.
}

\new{This list in Table~\ref{tab:pe_attributes} can be extended either by defining a new parameter in corresponding SoC file and parsing it or directly assigning as an attribute in \textit{PE class} of DS3 framework.}

\vspace{-10pt}
\subsection{Job Generator}
\label{ssec:job_generator}

\begin{figure}[b]
    \vspace{-10pt}
	\centering
	\includegraphics[width=0.9\linewidth]{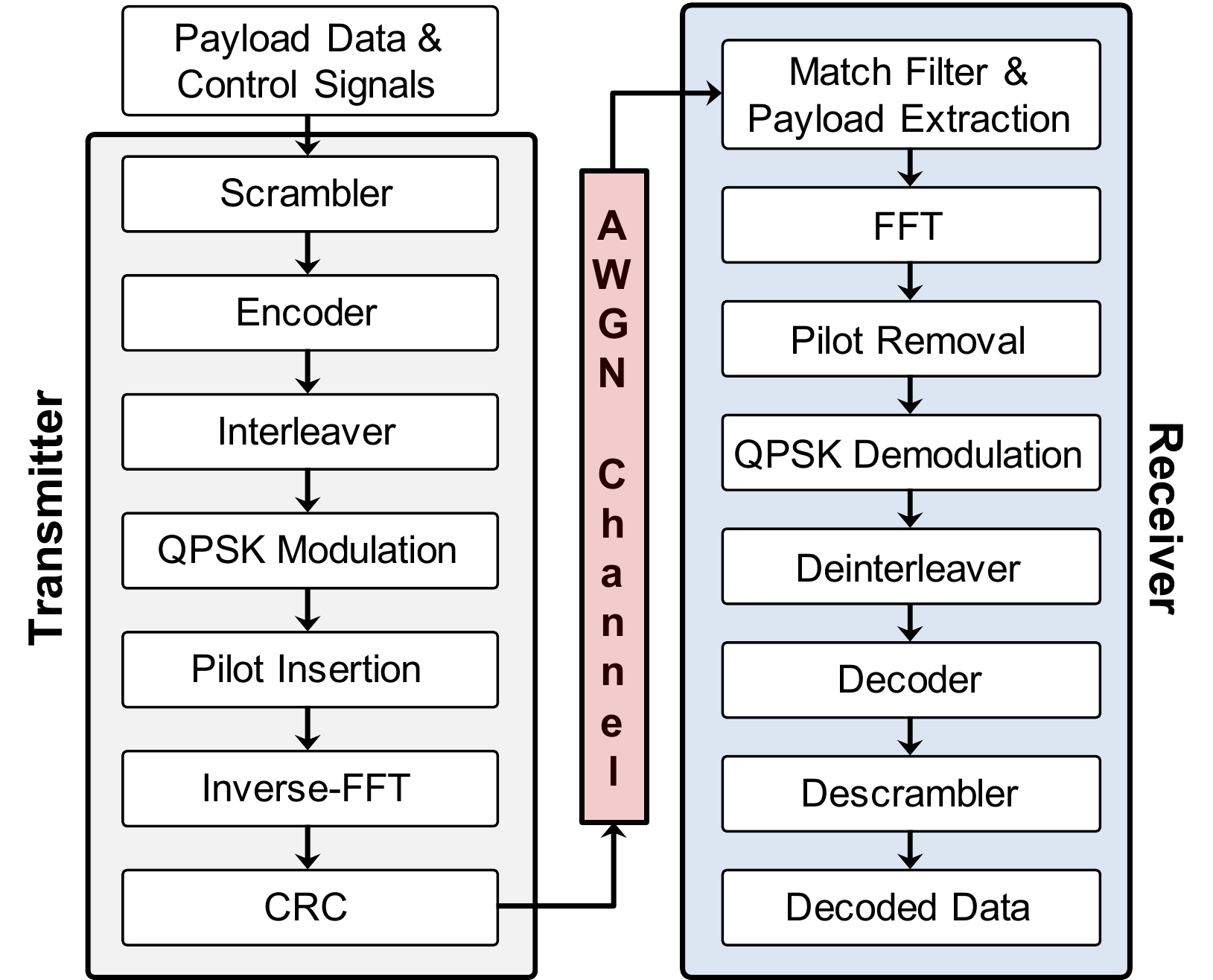}
	\caption{\new{Block diagrams for} WiFi-TX and WiFi-RX applications.}
	\label{fig:wifi_bd}
	\vspace{-15pt}
\end{figure}

Figure~\ref{fig:wifi_bd} presents block diagrams for a WiFi transmitter (WiFi-TX) and receiver (WiFi-RX) both of which are composed of multiple tasks. 
The tasks and dependencies in an application are represented using a DAG.
The job generator produces the tasks shown in Figure~\ref{fig:wifi_bd}, for a WiFi-TX job along with the dependencies \new{(see Appendix~\ref{appendix_DAGs} for real DAG implementations)}. 
The basic unit of data processed by this chain is a frame,
which is 64 bits in this work. 
DS3 defines each new input frame of an application as a job.
Hence, each job implies a 64-bit frame streaming through the WiFi-TX chain.

The job generator produces the tasks and DAG for each active application following a user-specified job injection model.
DS3 currently models the traffic by injecting jobs based on an exponential distribution.
The framework provides the flexibility to model other distributions as well. 
DS3 is scalable in terms of job generation and is capable of spawning jobs from multiple applications.
For example, suppose that both WiFi-TX and WiFi-RX applications are active and the corresponding injection ratio is 0.8:0.2. 
On an average, DS3 generates 4 WiFi-TX jobs for every WiFi-RX job.
This capability plays a crucial role in exploring multiple types of workloads in the domain, as demonstrated in Section~\ref{sec:case studies}.

\vspace{-10pt}
\subsection{Scheduling and DTPM Algorithms}

DS3 provides a plug-and-play interface to choose between different scheduling and DTPM algorithms.
Hence, developers can implement their own algorithms and easily integrate them with the framework.
To achieve this, developers define the new scheduling algorithm as a member function of the \textit{Scheduler} class. 
Then, the new scheduler is invoked from the \textit{run} method of the simulation core.
Scheduling algorithms vary significantly in their complexities and hence, require different inputs to map tasks to PEs.
To support this, DS3 provides a loosely defined interface to specify inputs to the schedulers as required.
The framework supports list schedulers (such as HEFT~\cite{Topcu} as well as table-based schedulers (such as integer linear programming), where schedule for all the tasks in a job is generated at the time of job injection.

DS3 provides built-in DTPM policies and facilitates the design of new DTPM algorithms. 
The policy is invoked periodically at every control epoch, 
which is parameterizable by the user. 
To minimize the run-time and power overhead of DTPM decisions, 
we use 10ms--100ms range following the common practice~\cite{bhat2018algorithmic}.
A policy of low complexity may use only the power state information of the PEs.
On the other hand, advanced algorithms may use PE utilization and more detailed performance metrics, such as number of memory accesses and retired instructions.
In addition, the DTPM policies have access to the resource management, including the power consumption and performance profiles therein. 
The decisions of the DTPM policy are evaluated and applied to the PEs at every control epoch.

Developers can add new scheduling and DTPM algorithms without modifying the rest of DS3. 
Modular design enables both maintaining existing interfaces and expanding them to support radically different algorithms.

\begin{figure}[b]
    \vspace{-10pt}
	\centering
	\includegraphics[width=1.0\linewidth]{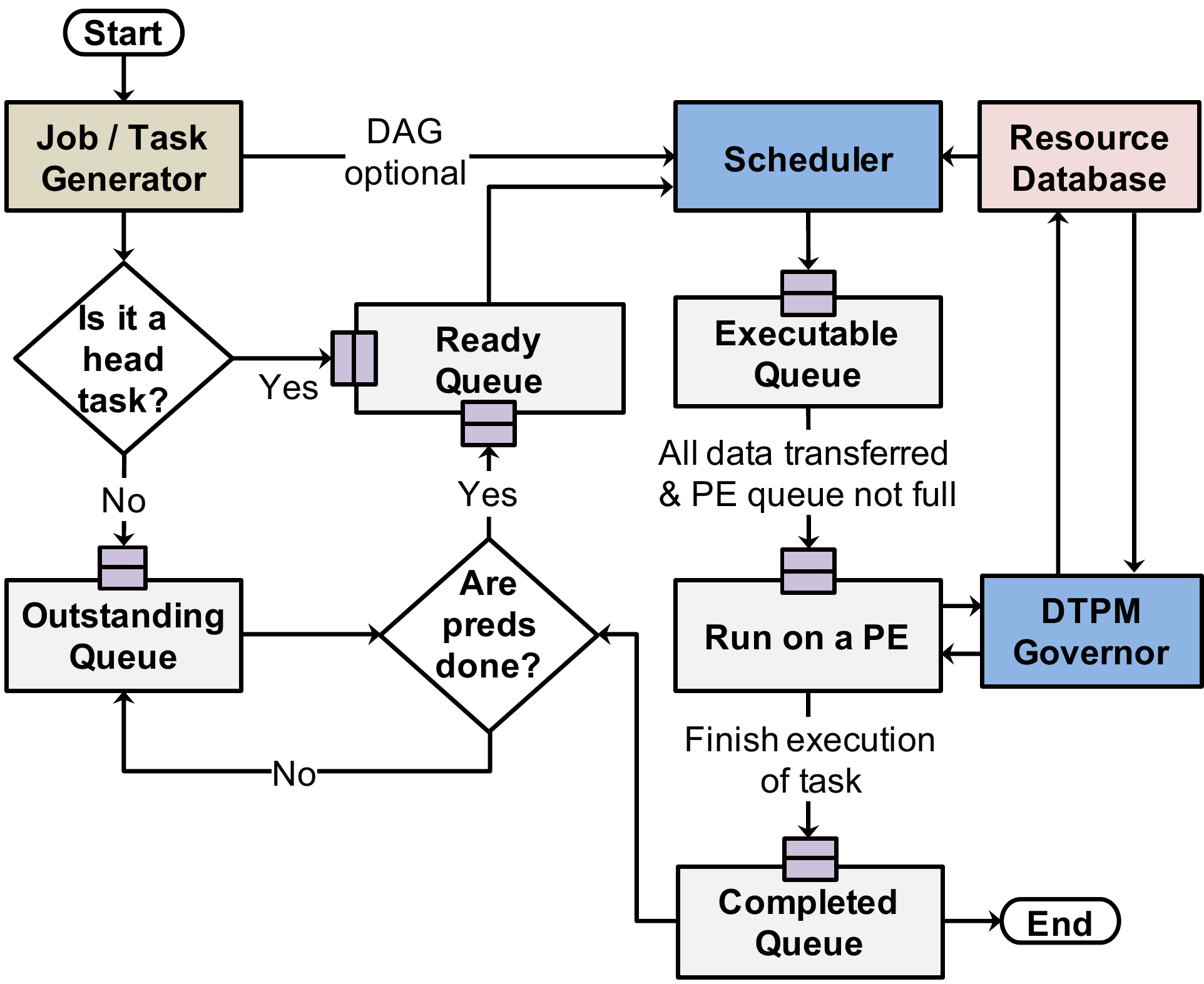}
	\caption{Life-cycle of a task in DS3 queues.}
	\label{fig:queues}
	\vspace{-10pt}
\end{figure}

\vspace{-10pt}
\subsection{Simulation Kernel}

The life cycle of a task in DS3 is shown in Figure~\ref{fig:queues}.
The job generator constructs a task graph as described in Section~\ref{ssec:job_generator}.
The tasks that are ready to execute (i.e. free of dependencies) are moved to a \textit{Ready Queue}. 
The other tasks that are waiting for predecessors to complete execution are held in the \textit{Outstanding Queue} before being moved to the \textit{Ready Queue}.
The scheduler\new{, an algorithm either built-in or user-defined, uses the resource database and produces PE assignments for ready tasks.}
Then, the simulation kernel migrates the tasks to the \textit{Executable Queue} until communication requirements from predecessors are met.
Finally, the task is simulated on the PE and retired after execution.
The simulation kernel clears the dependencies imposed by these tasks and removes them from the system. 
If all the predecessors of a task waiting in the \textit{Outstanding Queue} retire, 
then the kernel moves them to the \textit{Ready Queue}.
This triggers a new scheduling decision and the tasks experience a similar life cycle in the framework, as described above. 

\new{
Memory and network are shared resources in an SoC. 
The communication fabric to perform high-speed data transfer between the various resources in the platform is assumed to be a mesh-based network-on-chip (NoC). 
We integrate analytical models to compute the latency at a given traffic load in a priority-aware mesh-based industrial NoC~\cite{mandal2019analytical}. 
Executing multiple applications simultaneously leads to higher traffic in the network, as compared to the standalone execution. 
Hence, there is an effect in the execution time of applications if there is congestion in the network.
}

\begin{figure}[t]
	\centering
    	\includegraphics[width=1.0\linewidth]{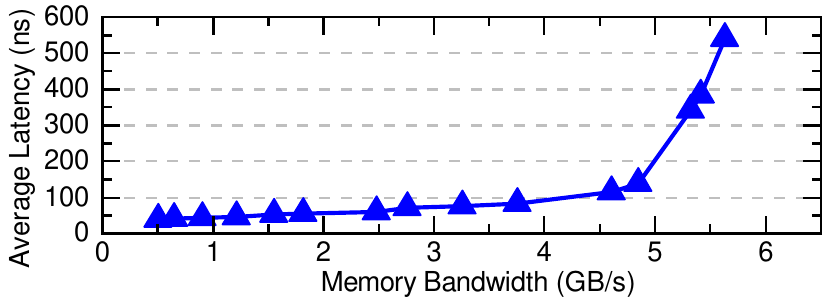}
	\caption{\new{Bandwidth-Latency curve used to model DRAM latency in DS3 framework.}}
	\label{fig:dram_latency}
	\vspace{-15pt}
\end{figure}

\new{
To model memory communication in the SoC, we include a bandwidth-latency model for memory latency modeling based on DRAMSim2~\cite{rosenfeld2011dramsim2}. 
DRAMSim2 is used to obtain memory latencies at varying bandwidth requirements as shown in Figure~\ref{fig:dram_latency}. 
DS3 models the transactions between the various communicating elements and keeps track of outstanding memory requests in a sliding window. 
We compute the memory bandwidth based on outstanding requests and then utilize the bandwidth-latency curve as a look-up table to obtain the average latency for the current memory bandwidth and add it to the execution time of the application(s). 
Hence, we account for contention of shared resources using the described network and memory models.
}

The simulation kernel also calls the DTPM governor periodically at every decision epoch.
The DTPM governor determines the power states of the PEs as a function of their current load and information provided by the resource database.
Subsequently, the simulation kernel updates the power states of the PEs and the decisions are retained until the next evaluation at the next epoch.

%% file: files/5-capabilities.tex
\vspace{-10pt}
\section{User View: DS3 Capabilities} \label{sec:capabilities}

This section presents the built-in scheduling and DTPM algorithms provided by the framework.

\vspace{-5pt}
\subsection{Scheduling Algorithms} \label{scheduler}

\begin{table}[t]
\vspace{-10pt}
\hspace{0.35cm}
\begin{minipage}[b]{0.5\linewidth}
\centering  
\includegraphics[width=0.9\linewidth]{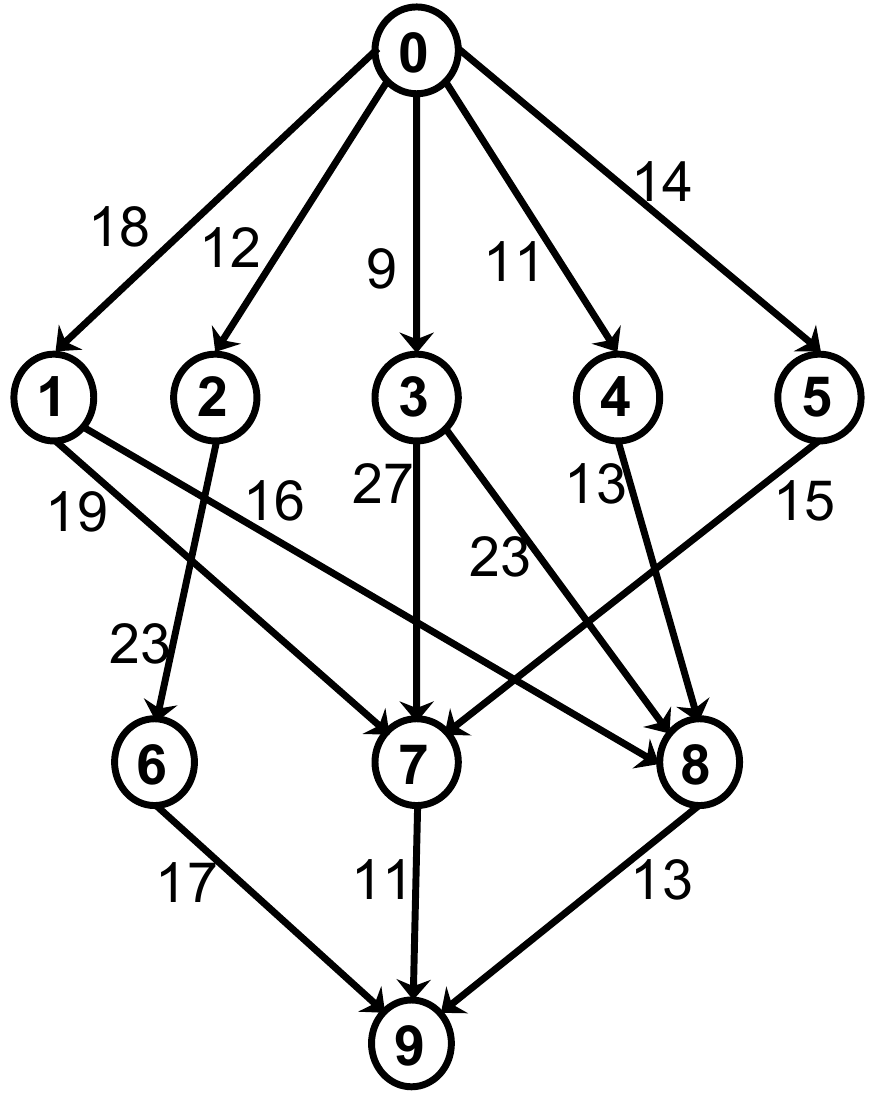}
\end{minipage}
\begin{minipage}[b]{0.5\linewidth}
\small
\begin{tabular}{@{}cccc@{}}
\toprule
\textbf{Task} & \textbf{P0} & \textbf{P1} & \textbf{P2} \\ \midrule
0 & 14 & 16 & 9 \\
1 & 13 & 19 & 18 \\
2 & 11 & 13 & 19 \\
3 & 13 & 8 & 17 \\
4 & 12 & 13 & 10 \\
5 & 13 & 16 & 9 \\
6 & 7 & 15 & 11 \\
7 & 5 & 11 & 14 \\
8 & 18 & 12 & 29 \\
9 & 21 & 7 & 16 \\ \bottomrule
\end{tabular}
\captionsetup{labelformat=empty}
\caption{}
\label{table:dag_execution_times}
\end{minipage}\hfill
\captionof{figure}{A canonical task flow graph~\cite{Topcu} with 10 tasks. Each node represents a task and each edge represents average communication cost across the available pool of PEs for the pair of nodes sharing that edge. The computation cost table on the right indicates the execution time for each of the PEs.}
\vspace{-15pt}
\label{fig:dag}
\end{table}

\begin{figure}[b]
    \vspace{-15pt}
	\centering
	\includegraphics[width=1\linewidth]{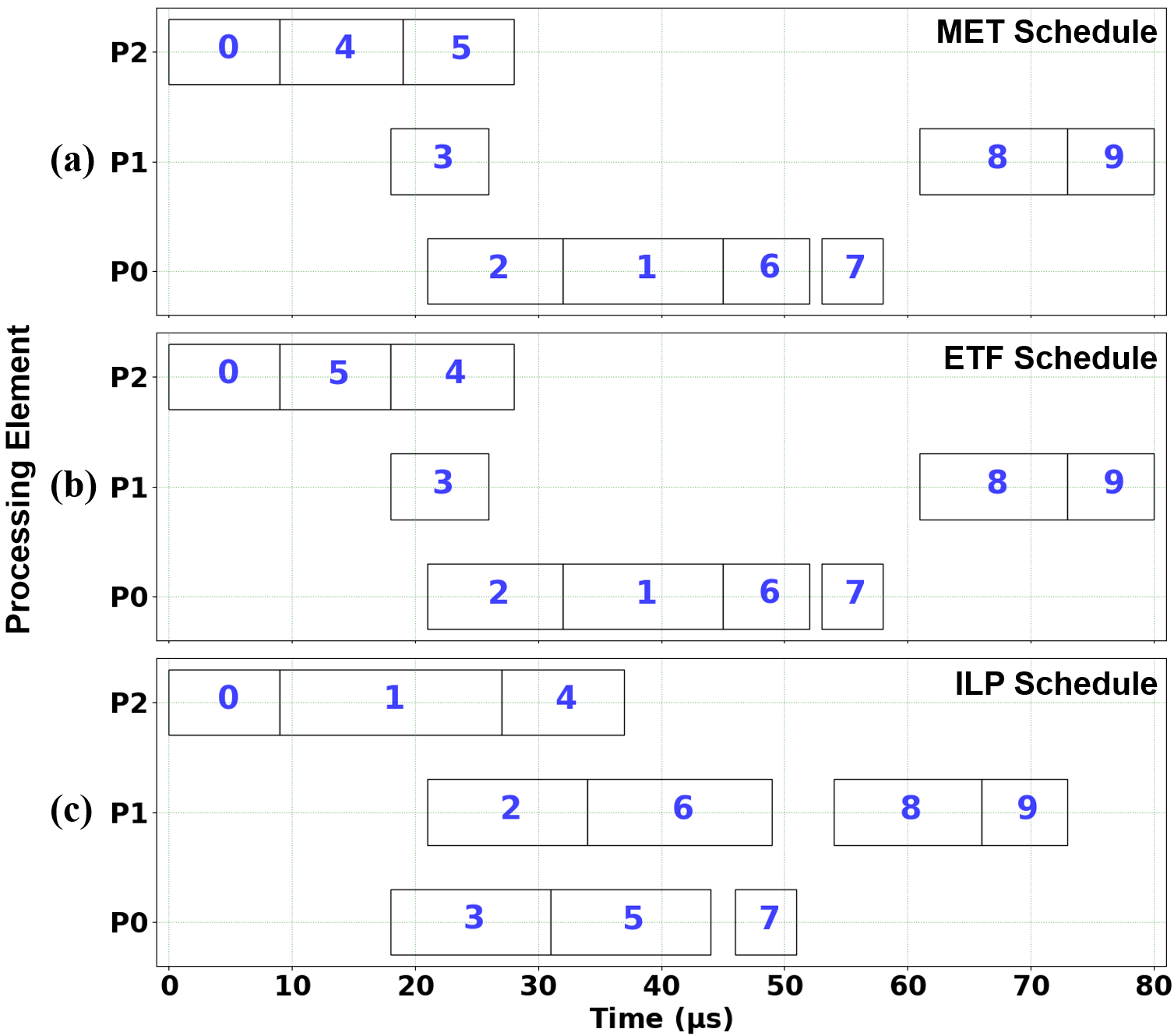}
	\caption{Schedule of task graph in Figure~\ref{fig:dag} with (a) MET, \\ (b) ETF, and (c) ILP.}
	\label{fig:schedule}
\end{figure}

DS3 provides a set of commonly used built-in scheduling algorithms, 
which can be specified by the users in the main configuration file. 
The framework generates Gantt charts to visualize the schedulers (see Figure~\ref{fig:schedule}).
This allows the end-user to understand the dynamics of the scheduler under evaluation.
We describe the built-in scheduling algorithms in DS3 using one of the most commonly used canonical task graphs~\cite{Topcu} shown in Figure~\ref{fig:dag}. 
Since this task graph is used commonly as a reference point for many list-scheduling studies, 
it serves as a representative example before analyzing the results from real-world applications in Section~\ref{ssec:case_studies}.

\vspace{3pt}
\noindent\textbf{Minimum Execution Time (MET) Scheduler: } 
The MET scheduler assigns a ready task to a PE that achieves the minimum expected execution time following a FIFO policy~\cite{heuristic_comparison}. 
If there are multiple PEs that satisfy the minimum execution criterion, the scheduler then reads the current state information of all these PEs and assign the tasks to one of the most idle PEs. 

Figure~\ref{fig:schedule}(a) shows the schedule generated by the MET scheduler in DS3 for the task graph shown in Figure~\ref{fig:dag}. All tasks are assigned to their best-performing PEs as expected.

\vspace{3pt}
\noindent\new{\textbf{Earliest Task First (ETF) Scheduler:}}
\new{The ETF scheduler utilizes the information about the communication cost between tasks and the current status of all PEs to make a scheduling decision~\cite{workflow_schedule}}. 

    




\new{Figure~\ref{fig:schedule}(b) shows the schedule produced by DS3 for a single instance of the task flow graph shown in Figure~\ref{fig:dag} with ETF scheduler.
Although the execution times are the same for both MET and ETF based schedules, the mapping decisions are different as shown in Figure~\ref{fig:schedule}(a) and ~\ref{fig:schedule}(b). 
This difference becomes evident when multiple applications are executed together, as shown in Section~\ref{ssec:case_studies}}.

\vspace{3pt}
\noindent\textbf{Table-based Scheduler:} 
DS3 also provides a scheduler which stores the scheduling decisions in a look-up table.  
This allows users to utilize any offline schedule, such as an assignment generated by an integer linear programming (ILP) solver, with the help of a small look-up table. 
For example, we use IBM ILOG CPLEX Optimization Studio~\cite{cplex} to generate the ILP solution for a job. 
The schedule from the ILP solution depicted in Figure~\ref{fig:schedule}(c) 
outperforms the other two schedulers for a single job instance as expected. 
However, we note that the schedule stored in the table guarantees optimality only if there is a single job in the system. 
Hence, its performance degrades when multiple jobs overlap, as shown 
in Section~\ref{ssec:case_studies}.


\def\bottomfraction{0.9}
\begin{figure*}
\centering
\begin{minipage}[c]{\textwidth}
\centering
	\includegraphics[width=1\linewidth]{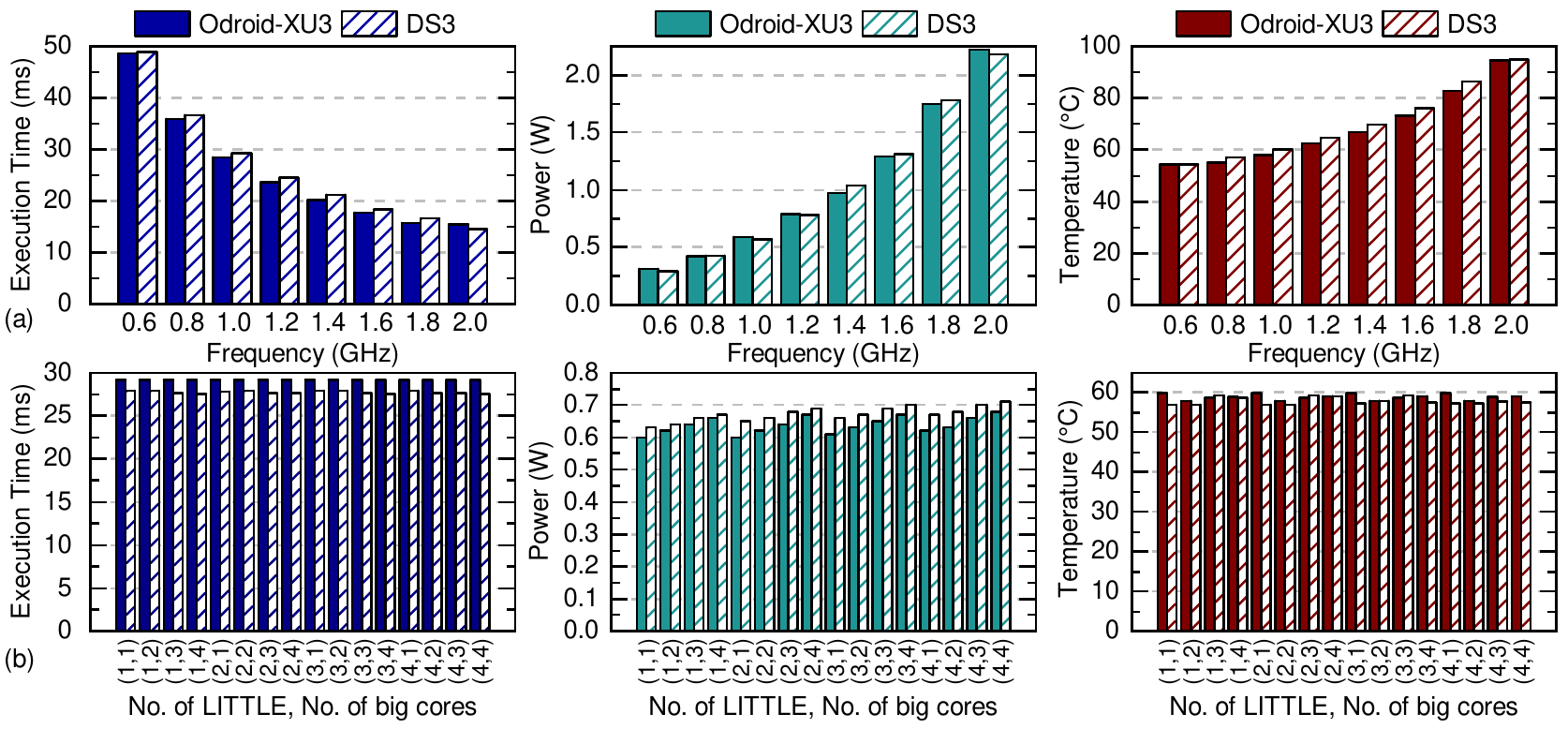}
	\caption{Comparison of execution time, power and temperature between DS3 and Odroid-XU3 for single-threaded applications when (a) Freq-Sweep: Number of cores is constant, frequencies of the cores are varied (b) Core-Sweep: Frequencies of cores are  constant, number of cores is varied.}
	\label{fig:validation_odroid_1}
\end{minipage}
\begin{minipage}[c]{\textwidth}
\centering
	\includegraphics[width=1\linewidth]{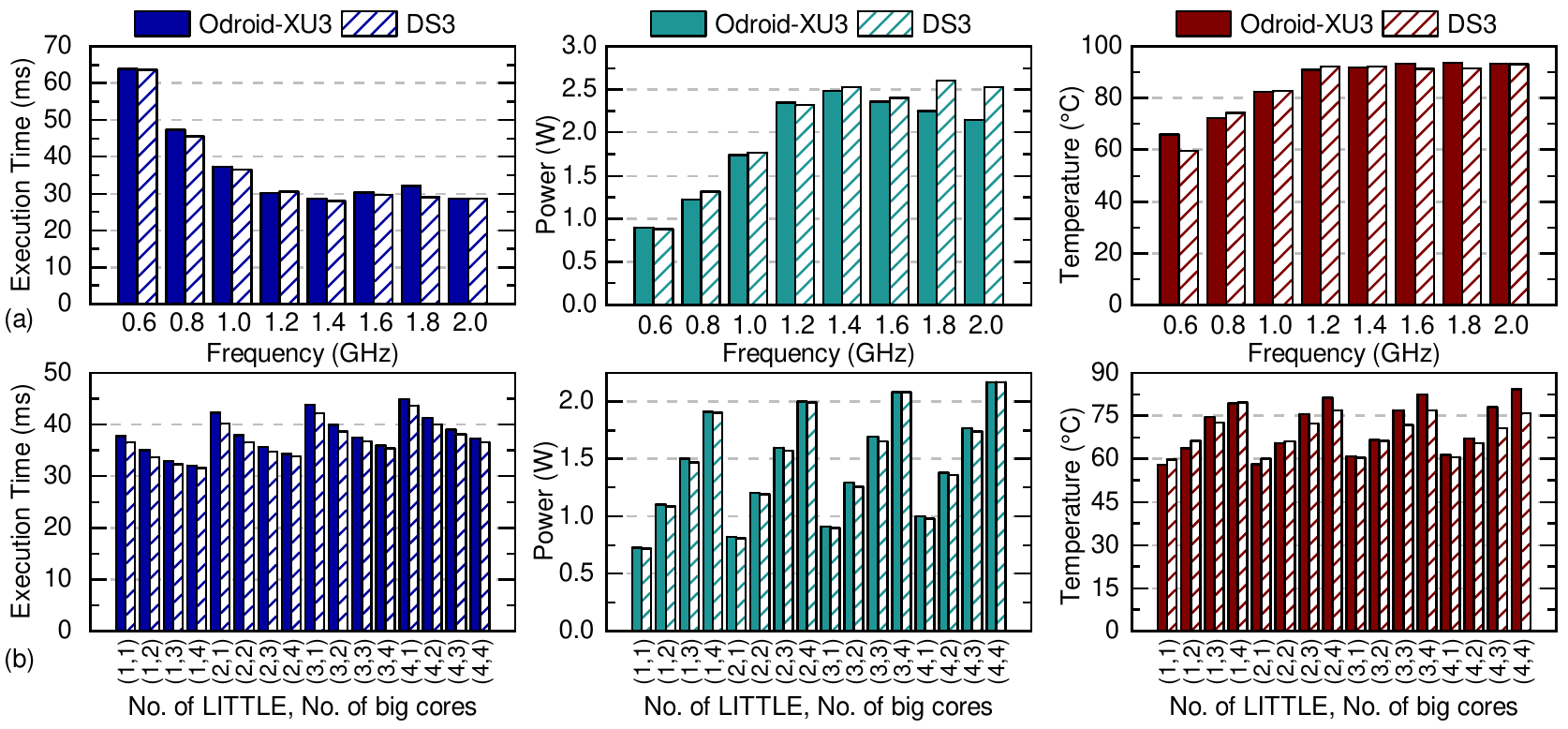}
	\caption{Comparison of execution time, power and temperature between DS3 and Odroid-XU3 for multi-threaded applications when (a) Freq-Sweep: Number of cores is constant, frequencies of the cores are varied (b) Core-Sweep: Frequencies of cores are  constant, number of cores is varied.}
	\label{fig:validation_odroid_2}
\end{minipage}
\begin{minipage}[c]{\textwidth}
\centering
	\includegraphics[width=1\linewidth]{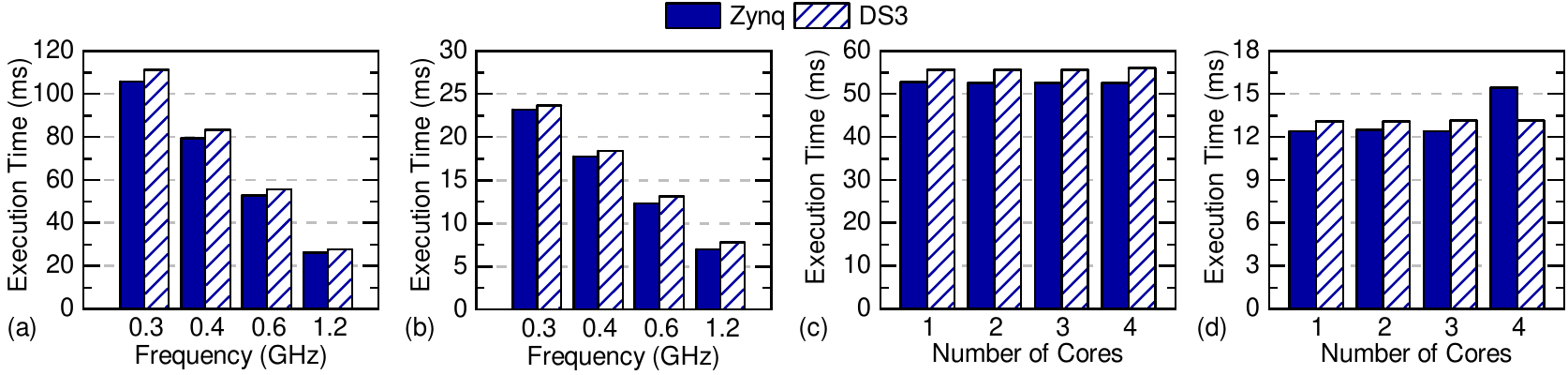}
	\caption{Comparison of execution time between DS3 and Zynq MPSoC for varying frequencies with (a) only Cortex A53 cores, (b) Cortex A53 cores and hardware accelerators and varying number of cores with (c) only Cortex A53 cores, (d) Cortex A53 cores and hardware accelerators.}
	\label{fig:validation_zynq}
\end{minipage}
\end{figure*}

\vspace{-10pt}
\subsection{DTPM Policies} \label{sec:dtpm}

State-of-the-art SoCs support multiple voltage-frequency domains and DVFS, which enables users to optimize for various power-performance trade-offs.
To support this capability, DS3 allows each PE to have a range of \textit{operating performance points} (OPPs), configurable in the resource database. 
The OPPs are voltage-frequency tuples that represent all supported frequencies of a given PE, 
which can be exploited by DTPM algorithms to tune the SoC at runtime.
In addition, DS3 integrates analytical power dissipation and thermal models for Arm Cortex-A15 and Cortex-A7 cores~\cite{bhat2018algorithmic}, as well as power consumption profiles for FFT~\cite{chen2018variable}, scrambler encoder, and Viterbi accelerators~\cite{tobola2018low}.

The power models capture both dynamic and static power consumption. 
The dynamic power consumption ($P=CV^2Af$) varies according to the load capacitance (\textit{C}), supply voltage (\textit{V}), activity factor (\textit{A}), and operating frequency (\textit{f}). 
Voltage and frequency are modeled through the OPPs, 
while the load capacitance and activity factors are modeled using measurements on real devices and published data. 
Static power consumption depends mainly on the current temperature and voltage, and DS3 uses thermal models obtained from measurements in the Odroid-XU3 SoC to accurately model both power and temperature.

The integrated power, performance, and temperature models enable us to implement a wide range of DTPM policies using DS3. 
To provide a solid baseline to the user, we also provide built-in DVFS policies that are commonly used in commercial SoCs. 
More specifically, users use the input configuration file to set the DTPM policy to \textit{ondemand, performance} and \textit{ powersave}~\cite{LinuxGovernors} governors, or to a custom DTPM governor.

\noindent \textbf{Ondemand Governor:} The \textit{ondemand} governor controls the OPP of each PE as a function of its utilization. 
The supported voltage-frequency pairs of a given PE are given by the following set: 
\begin{equation} \label{Vf_pairs}
    \mathcal{OPP} = \{(V_1, f_1), (V_2, f_2), \ldots, (V_k,f_k)\}
\end{equation}
where $k$ is the number of operating points supported by that PE. 
Suppose that the PE currently operates at $(V_2, f_2)$.
If the utilization of the PE is less than a user-defined threshold, then the \textit{ondemand} governor decreases the frequency and voltage such that the new OPP becomes $(V_1, f_1)$. 
If the utilization is greater than another user-defined threshold, the OPP is increased to the maximum frequency. 
Otherwise, the OPP stays at the current value, i.e., $(V_2, f_2)$.

\noindent \textbf{Performance Governor:} This policy sets the frequency and voltage of all PEs to their \textit{maximum} values to minimize execution time.

\noindent \textbf{Powersave Governor:} This policy sets the frequency and voltage of all PEs to their \textit{minimum} values to minimize power consumption.

\noindent \textbf{User-Specified Values:} This policy enables users to set the OPP (i.e., frequency and voltage) of each PE individually to a constant value within the permitted range. It enables thorough power-performance exploration by simply sweeping the supported values.
Finally, developers can also define custom DTPM algorithms in the \textit{DTPM} class, similar to the scheduler.

%% file: files/6-validation.tex
\vspace{-10pt}
\section{Simulator Validation} 
\label{sec:validation}

Simulation frameworks serve as powerful platforms to perform rapid design space exploration, evaluation of scheduling algorithms and DTPM techniques.
However, the fidelity of such simulation frameworks is questionable.
Particularly, the level of abstraction in high-level simulators is significant.
Hence, estimations from simulations may diverge from real SoCs due to differences in modeling, ineffective representation and limitations in simulation frameworks to capture overheads observed on hardware platforms.
In this section, we comprehensively evaluate our DS3 framework in terms of performance, power and temperature estimations with \new{two commercially available SoC platforms --- Odroid-XU3 and Zynq Ultracale+ ZCU-102.}

\vspace{-20pt}

\new{
\subsection{Validation with Odroid-XU3}
We choose Odroid-XU3 as one of the platforms for validation because of its abilities to measure power, performance and temperature.}
This platform comprises in-built current and temperature sensors enabling us to measure power, performance and temperature simultaneously and accurately at runtime.
Since the design space comprising the number of cores, frequency levels and applications is very large, we choose representative configurations and applications for validation against the Odroid-XU3, as shown in  Figure~\ref{fig:validation_odroid_1} and Figure~\ref{fig:validation_odroid_2}.
\new{For the comprehensive validation, we consider two cases: (1) Freq-Sweep: the number of cores is fixed, frequencies of the cores are varied and (2) Core-Sweep: the frequencies of the cores are fixed, the number of cores is varied.
To ensure completeness in validation, we validate both single-threaded and multi-threaded applications in both scenarios.
For single-threaded applications, Figure~\ref{fig:validation_odroid_1}(a) describes the execution time, power and temperature for Freq-Sweep scenario and Figure~\ref{fig:validation_odroid_1}(b) for Core-Sweep.
For Freq-Sweep, we vary the frequency from 0.6-2.0 GHz.
Odroid-XU3 has four LITTLE cores and four big cores, leading to 16 possible combinations of configurations of active cores.
Core-Sweep compares the parameters of DS3 and Odroid-XU3 for all 16 combinations.
Similarly, Figure~\ref{fig:validation_odroid_2}(a) and Figure~\ref{fig:validation_odroid_2}(b) show the comparisons of DS3 and Odroid-XU3 for multi-threaded applications.
Table~\ref{tab:validation_odroid} shows the error percentages in comparison of execution time, power and temperature for single-threaded and multi-threaded applications in both Freq-Sweep and Core-Sweep configurations.
}

\begin{table}[t]
\vspace{-5pt}
\centering
\caption{Error percentages in comparison of execution
time, power and temperature between DS3 and Odroid-XU3
}
\label{tab:validation_odroid}
\begin{tabular}{@{}ccccc@{}}
\toprule
\textbf{\begin{tabular}[c]{@{}c@{}}Application \\ Type\end{tabular}} & \textbf{Scenario} & \textbf{\begin{tabular}[c]{@{}c@{}}Execution \\ Time\end{tabular}} & \textbf{Power} & \textbf{Temperature} \\ \midrule
\multirow{2}{*}{\textbf{\begin{tabular}[c]{@{}c@{}}Single\\ Threaded\end{tabular}}} & Freq-Sweep & 3.6\% & 3.1\% & 2.7\% \\
 & Core-Sweep & 5.3\% & 5.1\% & 2.1\% \\ \midrule
\multirow{2}{*}{\textbf{\begin{tabular}[c]{@{}c@{}}Multi\\ Threaded\end{tabular}}} & Freq-Sweep & 2.8\% & 6.1\% & 2.4\% \\
 & Core-Sweep & 2.7\% & 1.3\% & 3.8\% \\ \bottomrule
\end{tabular}
\vspace{-10pt}
\end{table}

\new{In Freq-Sweep scenario, the frequency of the LITTLE and big cores are varied in unison until we reach the maximum frequency of the LITTLE cores, which is 1.4 GHz. 
Thereafter, the frequency of the big cores are swept until the maximum allowed frequency of 2.0 GHz.
We observe that the mean absolute error in performance, power and temperature estimates are \textit{3.6\%, 3.1\%, and 2.7\%}, respectively, for single-threaded applications.
Similarly, the errors are \textit{2.8\%, 6.1\% and 2.4\%} for multi-threaded applications.
}

\new{In Core-Sweep scenario, the number of active cores is varied in all combinations while the frequencies of the LITTLE and big cores are retained at 1.0 GHz.
The performance, power and temperature mean absolute errors are \textit{5.3\%, 5.1\%, and 2.1\%}, respectively, for single-threaded applications.
While multi-threaded applications have an average error of \textit{2.7\%, 1.3\% and 3.8\%}.
}

\new{
On an average, the error in accuracy is mostly less than 6\%.
We note that the platform experiences frequency throttling when the temperature reaches trip points (95\degree C).
The throttling behavior is modeled in DS3 and hence, we obtain highly correlated estimates for execution time, power and temperature even when the platform is throttled by the on-board thermal management agent.
In summary, the estimates from DS3 closely match real-time measurements obtained by the execution of similar workloads on the platform, as summarized in Table~\ref{tab:validation_odroid}.
The strong validation results aid in reinforcing the fidelity of the framework in simulating DSSoCs with high accuracy.
}

\vspace{-20pt}

\new{
\subsection{Validation with Zynq Ultrascale+ ZCU-102}
The second platform used to validate the results of the DS3 framework is Zynq Ultrascale+ ZCU-102 FPGA SoC. 
Zynq serves as a crucial platform for validation as it supports the implementation of hardware accelerators, unlike Odroid-XU3.
The support for hardware accelerators aids in validating DS3 against highly heterogeneous SoCs.
However, lack of on-board sensors prevent us from accurately measuring power and temperature. 
Hence, we chose to validate the execution time of Zynq and DS3 in the presence of hardware accelerators in various scenarios.
}

\new{We pick multiple scenarios to validate the execution time. 
First, we sweep the frequencies across the four supported frequencies on the Zynq board. 
We then measure the execution times when applications are executed only on Cortex A53 cores on the platform and then with both A53 cores and hardware accelerators.
Then, we vary the number of cores and measure the execution times with only A53 cores, and with A53 cores and hardware accelerators.
We observe high correlation between the measurements obtained from the Zynq board and DS3, as shown in Figure~\ref{fig:validation_zynq}.
However, when four A53 cores and hardware accelerators are enabled, we observe an anomaly with 15\% error between DS3 and Zynq. 
Current Linux kernels do not support scheduling and enablement of hardware accelerators in the operating system. Hence, we implement the scheduling mechanism for accelerators in user-space and identify the anomaly as an overhead that is incurred due to the user-space implementation. 
This overhead is expected to be significantly minimized when operating system kernels include support for accelerators. Finally, the average error in execution time is 6.85\%.}

%% file: files/7-case_studies.tex
\vspace{-12pt}
\section{Application Case Studies} \label{sec:case studies}

This section presents case studies and experiments for design space exploration of dynamic resource management, power-thermal management, and architecture configurations.
We base our studies on the benchmark applications in the domain of communications and radar, which are presented in the following section. 

\begin{figure}[b]
    \vspace{-15pt}
	\centering
	\includegraphics[width=1\linewidth]{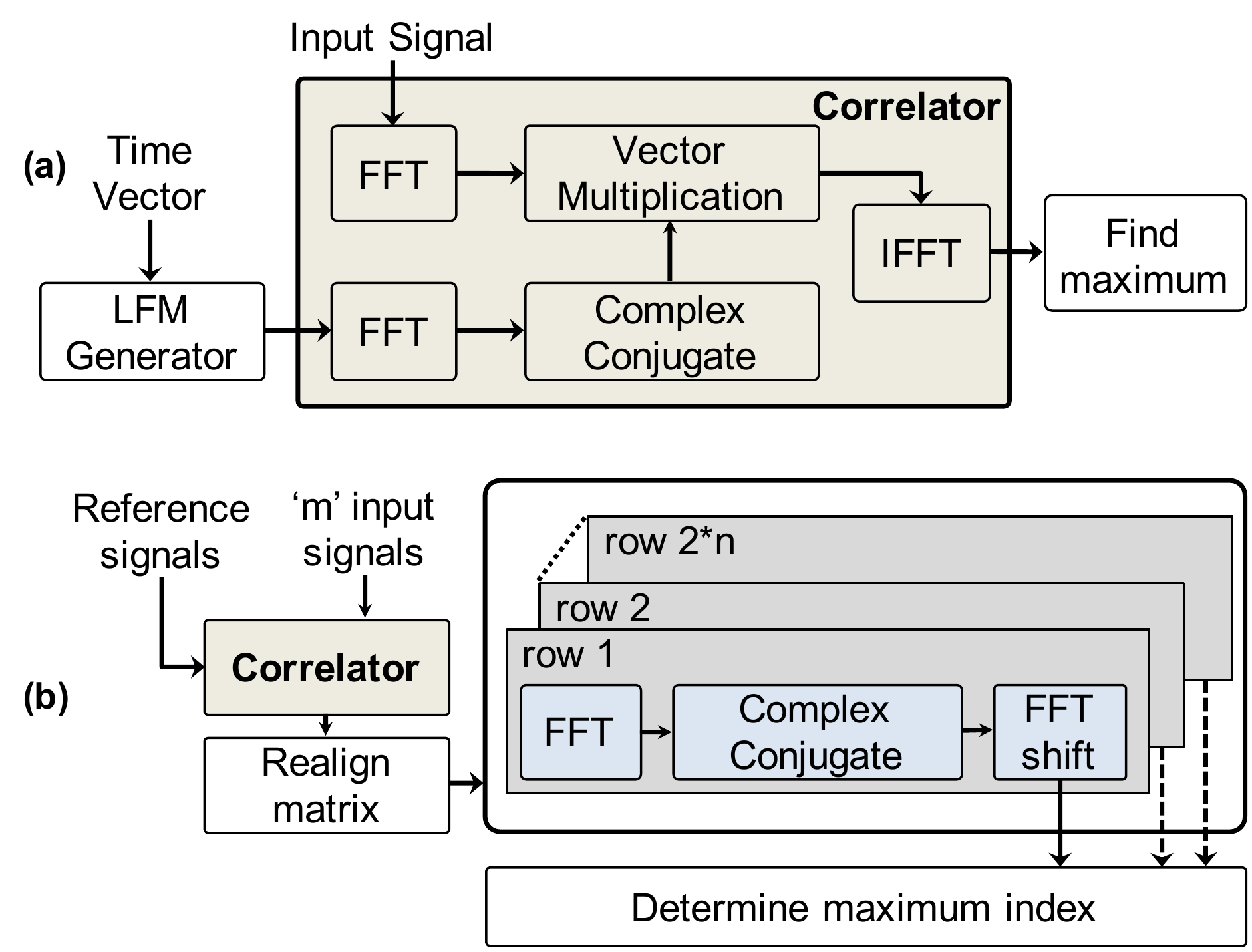}
	\caption{Block diagram of (a) range detection application, (b) pulse Doppler application where \emph{m} is number of signals and \emph{n} is number of samples for a signal.}
	\label{fig:radar_bd}
 	\vspace{-10pt}
\end{figure}

\begin{table}[t]
\centering
\small\addtolength{\tabcolsep}{-5pt}
\caption{\new{Execution profiles of applications on Arm A53 core in Xilinx ZCU-102, Arm A7/A15 cores in Odroid-XU3, and hardware accelerators}}
\label{tab:application_latencies}
\renewcommand{\arraystretch}{1}
\begin{tabular}{clcccc}
\hline
\toprule
\multirow{2}{*}{\textbf{Application}} & \multicolumn{1}{c}{\multirow{2}{*}{\textbf{Task}}} & \multicolumn{4}{c}{\textbf{Latency (\boldmath$\mu$s)}} \\ \cline{3-6} 
 & \multicolumn{1}{c}{} & \textbf{\begin{tabular}[c]{@{}c@{}}Zynq\\ A53\end{tabular}} & \textbf{\begin{tabular}[c]{@{}c@{}}Odroid\\ A7\end{tabular}} & \textbf{\begin{tabular}[c]{@{}c@{}}Odroid\\ A15\end{tabular}} & \textbf{\begin{tabular}[c]{@{}c@{}}HW\\ Acc.\end{tabular}} \\ \hline
\multirow{6}{*}{\textbf{\begin{tabular}[c]{@{}c@{}}WiFi\\ TX\end{tabular}}} & \begin{tabular}[c]{@{}l@{}}Scrambler-\\ Encoder\end{tabular} & 22 & 22 & 10 & 8 \\ \cline{2-6} 
 & Interleaver & 8 & 10 & 4 &  \\ \cline{2-6} 
 & \begin{tabular}[c]{@{}l@{}}QPSK\\ Modulation\end{tabular} & 15 & 15 & 8 &  \\ \cline{2-6} 
 & Pilot Insertion & 4 & 5 & 3 &  \\ \cline{2-6} 
 & Inverse-FFT & 225 & 296 & 118 & 16 \\ \cline{2-6} 
 & CRC & 5 & 5 & 3 &  \\ \hline
\multirow{8}{*}{\textbf{\begin{tabular}[c]{@{}c@{}}WiFi\\ RX\end{tabular}}} & Match Filter & 15 & 16 & 5 &  \\ \cline{2-6} 
 & \multicolumn{1}{c}{Payload Extraction} & 5 & 8 & 4 &  \\ \cline{2-6} 
 & FFT & 218 & 290 & 115 & 12 \\ \cline{2-6} 
 & Pilot Extraction & 4 & 5 & 3 &  \\ \cline{2-6} 
 & \begin{tabular}[c]{@{}l@{}}QPSK \\ Demodulation\end{tabular} & 79 & 191 & 95 &  \\ \cline{2-6} 
 & Deinterleaver & 10 & 16 & 9 &  \\ \cline{2-6} 
 & Decoder & 1983 & 1828 & 738 & 2 \\ \cline{2-6} 
 & Descrambler & 2 & 3 & 2 &  \\ \hline
\multirow{5}{*}{\textbf{\begin{tabular}[c]{@{}c@{}}Pulse\\ Doppler\end{tabular}}} & FFT & 30 & 35 & 15 & 6 \\ \cline{2-6} 
 & \begin{tabular}[c]{@{}l@{}}Vector\\ Multiplication\end{tabular} & 30 & 100 & 35 &  \\ \cline{2-6} 
 & Inverse-FFT & 30 & 35 & 15 & 6 \\ \cline{2-6} 
 & \begin{tabular}[c]{@{}l@{}}Amplitude\\ Computation\end{tabular} & 25 & 70 & 40 &  \\ \cline{2-6} 
 & FFT Shift & 6 & 7 & 3 &  \\ \hline
\multirow{5}{*}{\textbf{\begin{tabular}[c]{@{}c@{}}Range \\ Detection\end{tabular}}} & \begin{tabular}[c]{@{}l@{}}LFM Waveform\\ Generator\end{tabular} & 20 & 90 & 60 &  \\ \cline{2-6} 
 & FFT & 68 & 150 & 60 & 30 \\ \cline{2-6} 
 & \begin{tabular}[c]{@{}l@{}}Vector\\ Multiplication\end{tabular} & 52 & 75 & 60 &  \\ \cline{2-6} 
 & Inverse-FFT & 68 & 150 & 60 & 30 \\ \cline{2-6} 
 & Detection & 10 & 20 & 20 &  \\ \bottomrule
\end{tabular}
\vspace{-10pt}
\end{table}

\vspace{-10pt}
\subsection{Benchmark Applications} \label{ssec:benchmark}
\new{DS3 comes with \textit{six reference applications} from wireless communications and radar processing domain:}

\begin{itemize}
    \item \new{\textbf{WiFi-TX/RX},} 

    \item \new{\textbf{Low-power single-carrier TX/RX}, and}
    
    \item \new{\textbf{Radar and Pulse Doppler}.}
\end{itemize}

\new{The WiFi protocol consists of transmitter and receiver flows as shown in Figure~\ref{fig:wifi_bd}.
It has compute-intensive blocks, such as FFT, modulation, demodulation, and Viterbi decoder (see Table~\ref{tab:application_latencies}), which require a significant amount of system resources. When the bandwidth and latency requirements are small, one can use a simpler single carrier protocol to achieve lower power consumption. Finally, we include two applications from the radar domain as part of the benchmark application suite - (1) range detection and (2) pulse Doppler (see Table~\ref{tab:application_latencies}).
Figure~\ref{fig:radar_bd} represents block diagrams of the range detection and pulse Doppler algorithms.}

\new{The benchmark applications enable various algorithmic optimization and realistic design space exploration studies, as we demonstrate in this paper. Hence, we will continuously include applications from other domains to the benchmark suite.}

\vspace{-10pt}
\subsection{Reference Design of Applications}\label{ssec:reference_design}

\boldmath
\begin{figure*}[t]
	\centering
	\includegraphics[width=1\linewidth]{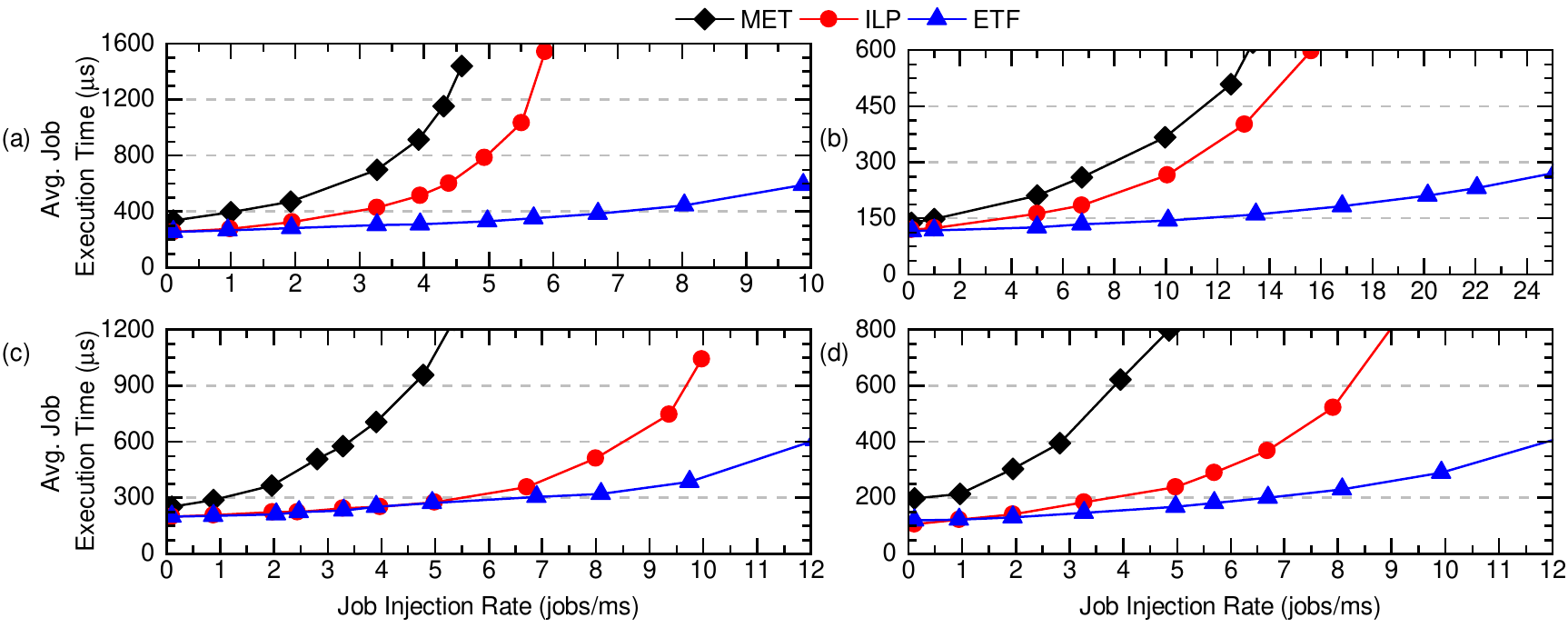}
	\caption{Results from different schedulers with a workload consisting of (a) WiFi-TX ($p_{TX}$=0.2) and WiFi-RX ($p_{RX}$=0.8), (b)  WiFi-TX ($p_{TX}$=0.8) and WiFi-RX ($p_{RX}$=0.2), (c) range detection ($p_{range}$=0.8) and pulse Doppler ($p_{pulse}$=0.2), (d) WiFi-TX ($p_{TX}$=0.3), WiFi-RX ($p_{RX}$=0.3), range detection ($p_{range}$=0.3), and pulse Doppler ($p_{pulse}$=0.1).}
	\label{fig:all_app}
	\vspace{-10pt}
\end{figure*}
\unboldmath

We developed a reference design for each of the applications described in Section~\ref{ssec:benchmark} on two popular 
commercial heterogeneous SoC platforms: 
Xilinx Zynq ZCU-102~UltraScale MpSoC~\cite{FPGA} and Odroid-XU3~\cite{ODROID} which has Samsung Exynos 5422 SoC. 
\new{
In the scope of DS3, we implement hardware accelerators in the programmable logic (PL) of the Xilinx Zynq SoC. 
Depending on the size of data transfers required for the accelerator, we use either memory-mapped (AXI4-Lite) or streaming interfaces (AXI-Stream). 
To be specific,  we use memory-mapped interfaces to communicate with scrambler-encoder accelerators and stream interfaces for the FFT accelerators. 
A direct memory access (DMA) unit facilitates data transfers between user-space mappable memory buffers and the accelerators using a streaming interface. 
In this regard, we profiled the computation and communication times in a Linux environment running on the Zynq SoC, which is then fed to the schedulers. 
The latency of each task in every application on different resource types is profiled, as shown in Table~\ref{tab:application_latencies}.
The schedulers decide to allocate the task to either hardware accelerators or general-purpose cores based on their corresponding function of communication- and computation-times along with other system parameters.}
In addition, we used the power consumption and temperature sensors on the Odroid-XU3 board for power consumption profiling.
DS3 power/performance models used in the resource manager incorporate these performance profiles for each task-resource pair.

\vspace{-10pt}
\subsection{Scheduler Case Studies}\label{ssec:case_studies}


This section provides an extension to our previous work in~\cite{arda2019simulation}, using 
built-in DS3 schedulers and applications in the benchmark suite. 
The simulations run on an SoC configuration that mimics a typical heterogeneous SoC with a total of 16 general purpose cores and hardware accelerators: 4 big Arm Cortex-A15 cores, 4 LITTLE Arm Cortex-A7, 2 scrambler accelerators, 4 FFT accelerators, and 2 Viterbi decoders.

\begin{table}[b]
\vspace{-10pt}
\centering
\caption{\label{tab:makespan}Execution time of applications in benchmark suite with different schedulers}
\begin{tabular}{lcccc}
\toprule
\multicolumn{5}{c}{\textbf{Execution Time of Single Job (\boldmath$\mu$s)}} \\ \hline
\multicolumn{1}{c}{} & \multicolumn{1}{l}{\textbf{WiFi-TX}} & \textbf{WiFi-RX} & \textbf{\begin{tabular}[c]{@{}c@{}}Range \\ Detection\end{tabular}} & \textbf{\begin{tabular}[c]{@{}c@{}}Pulse\\ Doppler\end{tabular}} \\ \hline
MET & 69 & 389 & 177 & 1665 \\
ETF & 69 & 301 & 177 & 1045 \\
ILP & 69 & 288 & 177 & 1000 \\ \bottomrule
\end{tabular}
\end{table}

We schedule and execute the WiFi TX/RX, range detection and pulse Doppler task flow graphs using DS3 and plot the average job execution time trend with respect to the job injection rate, as shown in Figure~\ref{fig:all_app}.
We use the parameters $p_{RX}$, $p_{TX}$, $p_{range}$, and $p_{pulse}$ representing the probabilities for the new job being WiFi-RX, WiFi-TX, range detection and pulse Doppler, respectively.

Figures~\ref{fig:all_app}(a) and (b) depict the results with WiFi applications for a download and upload intensive workload, independently.
To understand the performance of scheduling algorithms, we analyze the average execution time at varying rates of job injection.
MET uses a naive representation of the system state for scheduling decisions (described in Section~\ref{scheduler}), which results in higher execution time.
On the other hand, ILP uses a static table based schedule which is optimal for one job instance. 
At low injection rates (less than 1 job/$ms$), ILP is suitable as jobs do not interleave.
However, as the injection rate increases, the ILP schedule is not optimal. 
The ETF performance is superior in comparison to the others, as observed in Figures~\ref{fig:all_app}(a) and (b).

Figure~\ref{fig:all_app}(c) demonstrates the results for a workload comprising radar benchmarks.
This workload uses $p_{range}$ = 0.8 and $p_{pulse}$ = 0.2, owing to the difference in execution times of the two applications.
The performance of ETF and ILP schedulers are similar until 5 jobs/$ms$, following which performance of ETF is superior in comparison to ILP. 
Although the trend in execution time for radar benchmarks is similar to WiFi, the job injection rate at which ETF and ILP diverge is different because of the differences in execution times of these applications, as shown in Table~\ref{tab:makespan}.
At an injection rate lower than 5 jobs/$ms$, the level of interleaving of jobs is low which aligns with the ILP solution.

Finally, we construct a workload comprising of all four applications and Figure~\ref{fig:all_app}(d) shows the corresponding results.  
The performance trend of the schedulers with all applications is similar to WiFi and radar workloads.
MET considers only the best performing PEs for mapping and ILP is sub-optimal at high injection rates whereas ETF utilizes the state information of all PEs for mapping decisions.

In summary, the experiments presented in Figure~\ref{fig:all_app} demonstrate the capabilities of the simulation environment. DS3 allows the end user to evaluate workload scenarios exhaustively by sweeping the $p_{TX}$, $p_{RX}$, $p_{range}$ and $p_{pulse}$ configuration space to determine the scheduling algorithm that is most suitable for a given SoC architecture and set of workload scenarios.

\vspace{-10pt}
\subsection{SoC Design Space Exploration} \label{ssec:SoC_DSE}

\begin{table}[t]
\centering
\caption{\label{tab:SoC_DSE}Area, performance and energy for different SoC configurations with varying number of accelerators}
\begin{tabular}{@{}cccccc@{}}
\toprule
\multicolumn{3}{c}{\textbf{Configuration}} & \multirow{2}{*}{\textbf{\begin{tabular}[c]{@{}c@{}}Area\\ (mm$^2$)\end{tabular}}} & \multirow{2}{*}{\textbf{\begin{tabular}[c]{@{}c@{}}Average Job \\ Execution (\boldmath$\mu$s)\end{tabular}}} & \multirow{2}{*}{\textbf{\begin{tabular}[c]{@{}c@{}}Energy per \\ Job (\boldmath$\mu$J/job)\end{tabular}}} \\ \cmidrule(r){1-3}
ID & FFT & Viterbi &  &  &  \\ \midrule
1 & 0 & 0 & 14.94 & 2606 & 1744 \\
2 & 0 & 1 & 14.94 & 1824 & 1244 \\
3 & 2 & 1 & 15.82 & 293 & 589 \\
4 & 4 & 0 & 16.29 & 1212 & 957 \\
5 & 4 & 1 & 16.56 & 274 & 584 \\
6 & 6 & 3 & 19.29 & 264 & 582 \\ \bottomrule
\end{tabular}
\vspace{-10pt}
\end{table}

This section illustrates how DS3 can be utilized to identify the number and types of PEs during early design space exploration.
We employ the WiFi-TX and WiFi-RX applications to explore different SoC architectures. 
All configurations in this study have 4 big Arm Cortex-A15 and 4 LITTLE Arm Cortex-A7 cores to start with and DS3 guides the user to determine the number of configurable hardware accelerators in the architecture. 
We choose accelerators for FFT and Viterbi decoder. FFT is a widely used accelerator among all the applications, whereas Viterbi decoder execution cost on a general-purpose core is significantly high.

\subsubsection{\new{Grid Search}} \label{sssec:grid}
We vary the number of instances of FFT (0, 1, 2, 4, 6) and Viterbi decoder (0, 1, 2, 3). Table~\ref{tab:SoC_DSE} lists the representative configurations used in \new{this grid search} study. 
Each row in the table represents the configuration under investigation with an estimated SoC area, and average execution time and average energy consumption per job.
The DS3 framework provides metrics that aid the user in choosing a configuration that best suits power, performance, area and energy targets.

\begin{figure}[t]
	\centering
    	\includegraphics[width=1.0\linewidth]{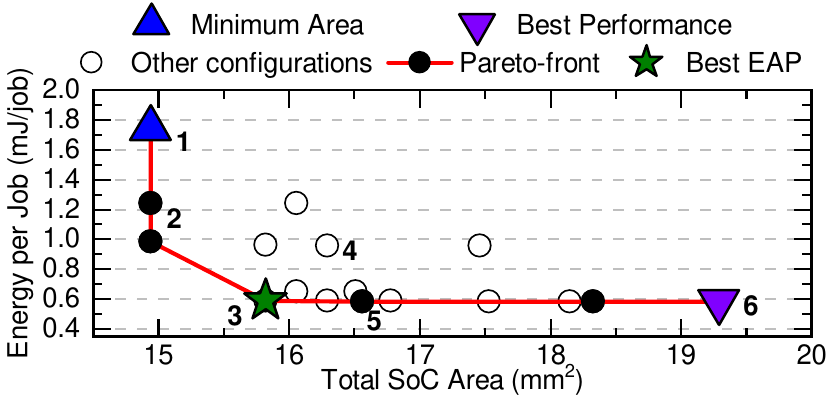}
	\caption{Design space exploration studies showing energy per job vs. SoC area with pareto-frontier.}
	\label{fig:SoC_DSE}
	\vspace{-10pt}
\end{figure}

\renewcommand{\thefigure}{15}
\begin{figure*}[h]
	\centering
    	\includegraphics[width=1\linewidth]{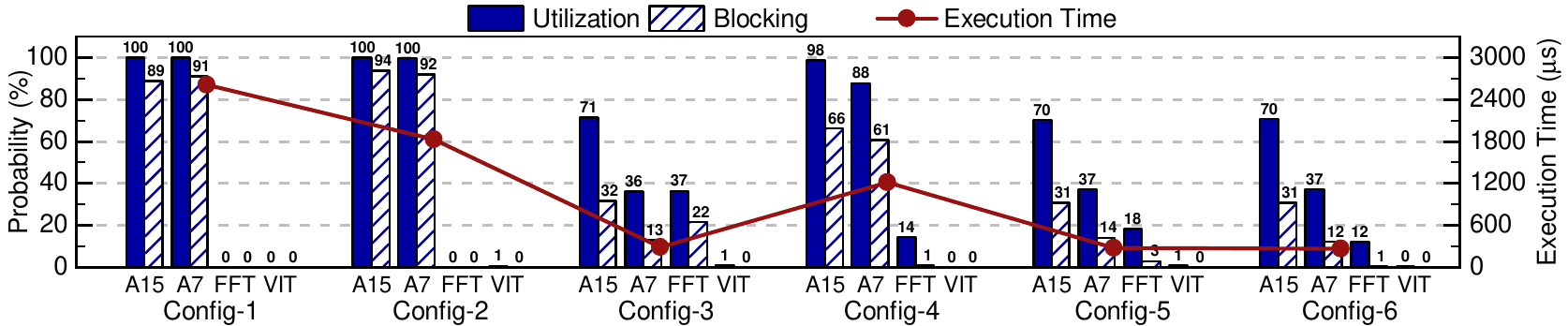}
	\caption{\new{Utilization vs blocking for PE clusters in representative configurations with average job execution times.}}
	\label{fig:Block}
	\vspace{-5pt}
\end{figure*}

\renewcommand{\thefigure}{14}
\begin{figure}[t]
    \vspace{-10pt}
	\centering
    	\includegraphics[width=0.6\linewidth]{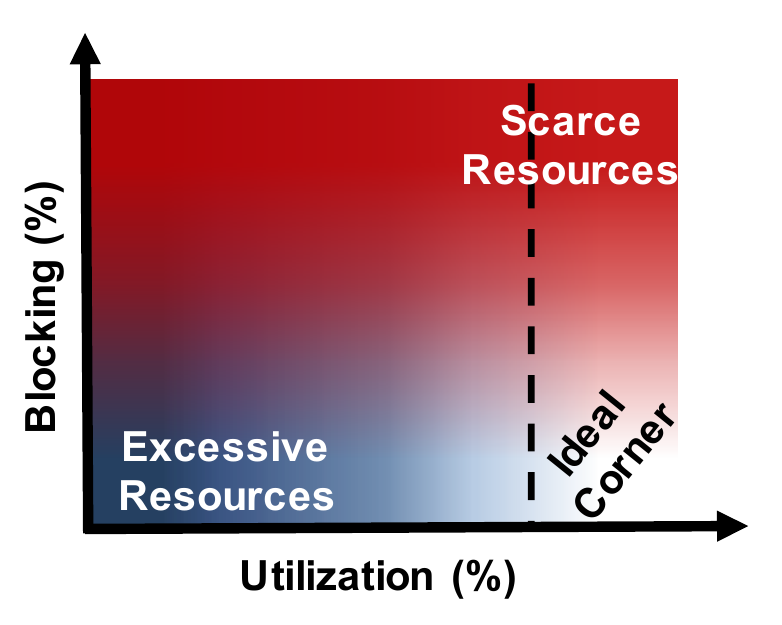}
	\vspace{-5pt}
	\caption{\new{PE blocking vs utilization (2-D performance plane)}}
	\label{fig:perf_plane}
	\vspace{-10pt}
\end{figure}

Figure~\ref{fig:SoC_DSE} plots the energy consumption per job as a function of the SoC area. 
We find the area of a given configuration using a built-in floorplanner that takes the areas of PEs and other components such as system level cache and memory controllers. 
The energy consumption per job is computed by dividing total energy consumption of the SoC for the entire workload with the number of completed jobs.

As the accelerator count increases in the system, 
the energy consumption per operation decreases. 
This comes at the cost of larger SoC area, as shown in Figure~\ref{fig:SoC_DSE} and Table~\ref{tab:SoC_DSE}. 
For this workload, \textit{configuration-3}, i.e., an SoC with two FFT and one Viterbi decoder accelerators, provides the best trade-off. 
Removing any of the accelerators leads to a significant increase in energy per operation with a small area advantage. 
In contrast, any further increase in the number of accelerators does not result in significant improvement in energy per job for this workload. 
As a result, \textit{configuration-3} is the best configuration in terms of energy-area product (EAP). 
This configuration leads to an EAP gain of almost 65\% (an energy reduction of 67\% with an increase in area by only less than 6\%) compared to \textit{configuration-1}.
After this point, the improvement in the overall performance by adding more FFT accelerators and Viterbi decoders does not overcome the cost of increase in the total area as seen in both Figure~\ref{fig:SoC_DSE} and Table~\ref{tab:SoC_DSE}.

\subsubsection{\new{Guided Search}} \label{sssec:guided}

\new{DS3 also supports a guided search in the design space. Figure~\ref{fig:perf_plane} depicts a 2-D performance plane for a PE (or a PE cluster) where \textit{x}-axis and \textit{y}-axis are utilization and blocking, both in percent, respectively. Ideally, a PE should be on the lower-right corner where utilization is high, and blocking is low. If both utilization and blocking are high, upper-right corner, then it means that there is a need for more resources in the system. If, however, the opposite is true, the utilization of a PE is low, and it also does not block tasks very often. In this case, resources in the system are abundant. Finally, a PE should never be on the upper-left corner, representing low utilization and high blocking which is unrealistic.}

\new{Considering the case study in Section~\ref{sssec:grid} where we explored 20 different configurations, the guided search will converge on the best configuration faster.  Figure~\ref{fig:Block} shows how utilization, blocking, and average job execution time differ for six aforementioned configurations. \textit{Configuration-1}, with no FFT and Viterbi accelerator, yields a high utilization and blocking for both Arm clusters, hence SoC requires hardware accelerators. The results with \textit{configuration-2}, addition of one Viterbi accelerator,  indicate that Viterbi accelerator is a critical component for the system since it provides a huge gain in average job execution time (a reduction from 2606 $\mu$s to 1824 $\mu$s) although the utilization for this accelerator is very small (0.61\%). \textit{Configuration-2} also suggests that one Viterbi accelerator is enough for the system since both utilization and blocking is low. Based on this observation, we directly eliminate configurations with no and more than one Viterbi accelerator (i.e., \textit{configuration-4} and \textit{-6}, see Table~\ref{tab:SoC_DSE}). The comparison between \textit{configuration-3} and \textit{-5} based on utilization, blocking, and estimated SoC area draws a conclusion that \textit{configuration-3} is the best configuration for this case study.}

\setcounter{figure}{15}
\renewcommand{\thefigure}{\arabic{figure}}
\begin{figure}[tb]
	\centering
    	\includegraphics[width=1.0\linewidth]{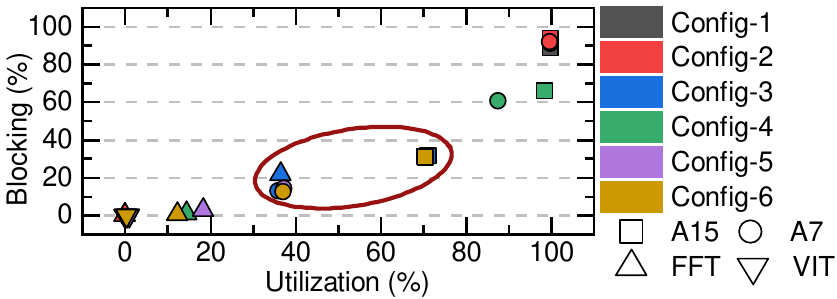}
	\vspace{-5pt}
	\caption{\new{Results for representative configurations on 2-D performance plane for PE clusters.}}
	\label{fig:perf_plane_results}
	\vspace{-10pt}
\end{figure}

\new{Figure~\ref{fig:perf_plane_results} depicts the same results for the representative configurations on the 2-D plane. As seen, \textit{configuration-3} is the closest one to the ideal region and provides the best trade-off.}

\new{These approaches} can be used to utilize DS3 for design space exploration of SoC architectures by analyzing energy efficiency, area, and developmental effort involved in the development of specialized cores for hardware acceleration. This approach can also be extended to explore the effect of increase (decrease) in the number of general-purpose cores and graphics processing units (GPUs) in a DSSoC architecture.

\vspace{-10pt}
\subsection{DTPM Design Space Exploration} \label{sec:DTPM}

In this section, we evaluate a subset of five applications from our benchmark set: WiFi-RX/TX, single-carrier RX/TX, and range detection on an SoC with 16 heterogeneous PEs. We explore 8 frequency points for the big cluster (0.6-2.0GHz) and 5 for the LITTLE (0.6-1.4GHz), using a 200MHz step. All possible DVFS modes were evaluated, i.e., all possible combinations of power states for each PE in addition to \textit{ondemand}, \textit{powersave}, and \textit{performance}~\cite{LinuxGovernors} modes.

\begin{figure}[t]
    \vspace{-15pt}
	\centering
	\includegraphics[width=1.0\linewidth]{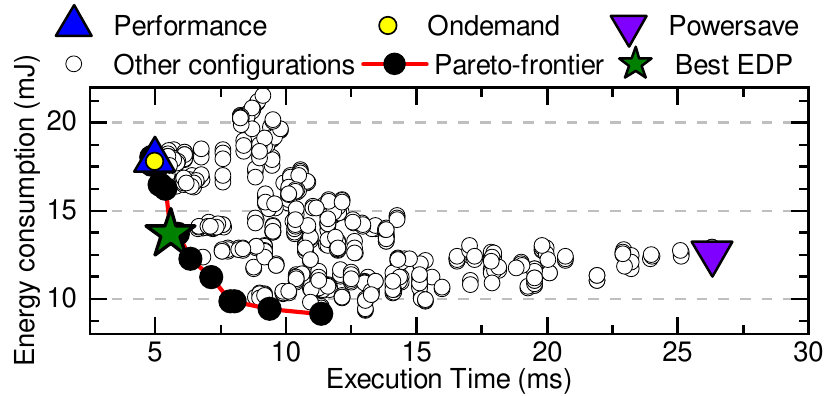}
	\caption{Pareto frontier in the energy-performance curve for an SoC with 16 processing elements (PEs) executing a representative workload.}
	\label{pareto-front}
	\vspace{-10pt}
\end{figure}

Figure~\ref{pareto-front} presents the Pareto frontier for all aforementioned configurations. The \textit{ondemand} and \textit{performance} policies provide low latency with high energy consumption, while \textit{powersave} minimizes the power at the cost of high latency, which results in sub-optimal energy consumption due to the increased execution time. The best configuration in terms of EDP uses 1.6GHz and 4 active cores for the big cluster, and 600MHz and 3 active cores for the LITTLE cluster, achieving 5.6ms and 13.7mJ. This figure also shows a $5\times$ variation in execution time and energy consumption between different configurations. The best EDP configuration achieves up to $4\times$ better EDP than default DTPM algorithms. This indicates that there is opportunity for users to propose their own power management mechanisms to further improve the energy efficiency of the system and integrate those mechanisms into DS3.

\begin{figure}[b]
    \vspace{-10pt}
	\centering
	\includegraphics[width=1.0\linewidth]{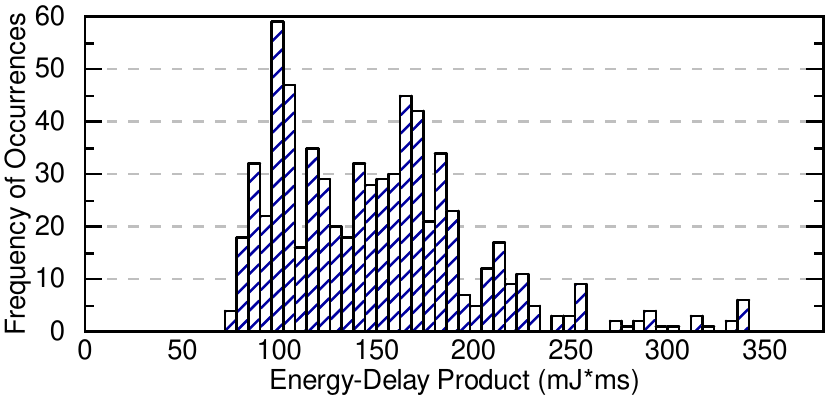}
	\caption{Energy-Delay Product (EDP) histogram.}
	\label{histogram-EDP}
	\vspace{-15pt}
\end{figure}

In addition to the Pareto frontier, DS3 also provides energy, performance, and EDP histograms to aid the design space exploration. Due to space constraints, Figure~\ref{histogram-EDP} depicts only the EDP histogram.
This histogram shows that only a small fraction of configurations achieve an EDP below 80mJ*ms. While the \textit{performance} and \textit{ondemand} governors are in the range of 90mJ*ms, the \textit{powersave} mode gets 330mJ*ms EDP. 
Therefore, users can use these visualization tools to quickly identify the most promising configurations for the SoC.

\vspace{-10pt}
\subsection{Scalability Analysis} \label{sec:scalability}

\begin{figure}[t]
	\centering
	\includegraphics[width=1.0\linewidth]{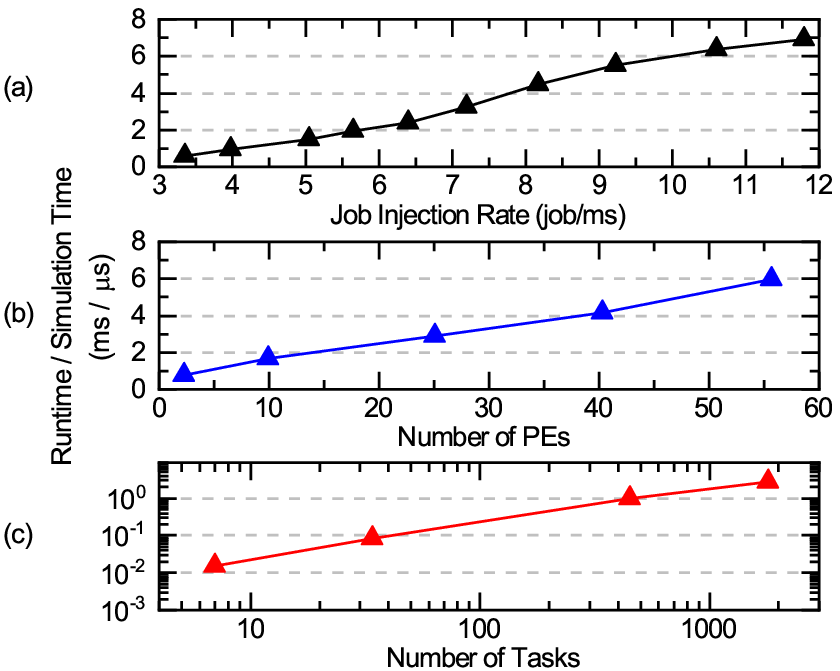}
	\caption{Results for scalability analysis showing DS3 runtime versus (a) Different job injection rates, (b) Varying SoC configurations and (c) Number of tasks executed.}
	\label{fig:scale}
 	\vspace{-10pt}
\end{figure}

\begin{figure*}[t]
    \vspace{-10pt}
	\centering
	\includegraphics[width=1.0\linewidth]{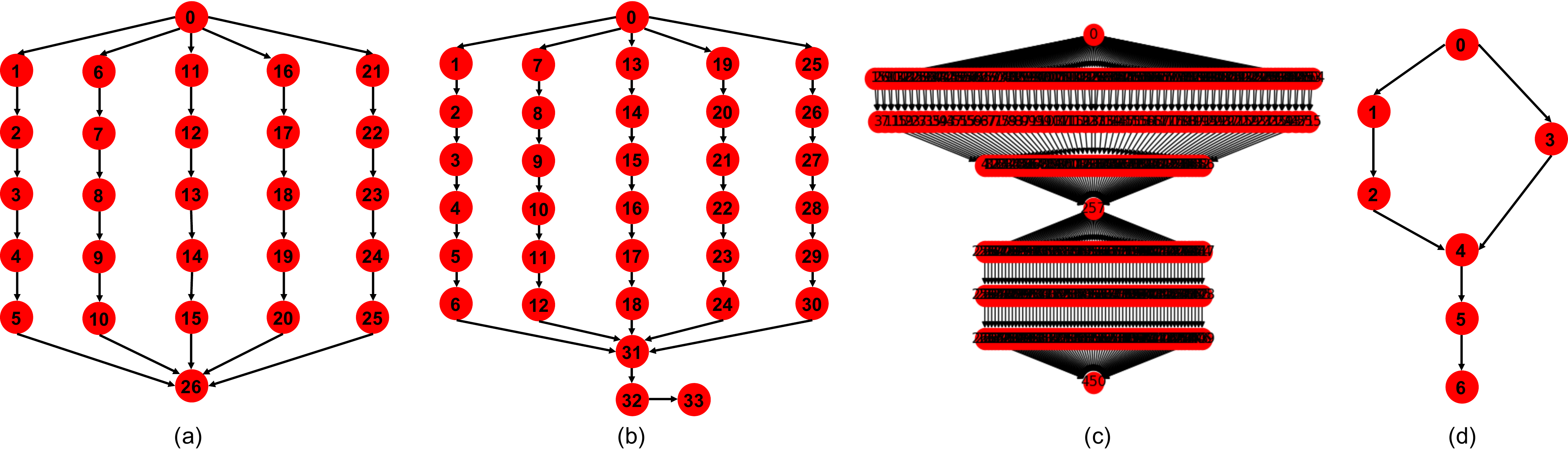}
	\caption{\new{DAG representations for (a) WiFi-TX, (b) WiFi-RX, (c) pulse Doppler, and (d) range detection application.}}
	\label{fig:dag_rep}
	\vspace{-10pt}
\end{figure*}

This section illustrates the scalability of DS3 as a function of simulated number of jobs, SoC size, and the number of tasks in a single job (application).
To maximize the load, we run four applications (WiFi TX/RX, Range detection and Pulse Doppler) simultaneously 
while sweeping the job injection rates and number of PEs. For the last case, however, we fix the job injection rates and SoC configuration while running applications different in size, separately. 

Figure~\ref{fig:scale}(a) shows the total simulation time as a function of the number of jobs injected throughout the simulation. As the relation between two metrics is linear, DS3 simulation run-time increases linearly with the workload size.
Similarly, Figure~\ref{fig:scale}(b) presents the simulation time when the number of PEs increase. 
This relationship is also linear, leading to 6ms per simulation cycle (1$\mu$s) for a 56-core configuration.
Finally, in Figure~\ref{fig:scale}(c), the simulation time with respect to application size (number of tasks in a single job) is depicted. As application size grows, the runtime to simulate of 1$\mu$s also increases linearly.

The scalability analysis provided in this section demonstrates that DS3 runtime is a linear function of workload, SoC and application size. 
As a side note, we obtain a simulation speedup of $~600\times$ when running 1675 jobs of WiFi-TX in comparison to gem5. Hence, DS3 facilitates rapid design space exploration with relatively short turn-around times.
This feature is important as DS3 aims to help users with extensive design space exploration in a relatively short period while avoiding unnecessary simulation of low-level details.

%% file: files/8-conclusion.tex
\vspace{-10pt}
\section{Conclusion and Future Work}\label{sec:concandfuture}

The performance and energy efficiency potential of DSSoCs remains untapped unless we employ efficient run-time resource management techniques. 
This paper presented DS3, a Python-based open-source system-level framework for rapid design space exploration of domain specific SoCs. 
We developed a scalable, modular, and flexible simulation framework to evaluate scheduling algorithms, dynamic power-thermal management techniques and architecture exploration. 
We also presented benchmark applications, built-in scheduling algorithms and DVFS policies which can be used as reference by users and developers.
Finally, the framework is thoroughly validated against a commercially available SoC, asserting the fidelity of the simulator to successfully simulate domain-specific SoCs.

We shared DS3, built-in schedulers, DVFS governors, and benchmark applications to the public to encourage research in the domain of DSSoCs.
The integration of learning-based schedulers and expanding the domain beyond wireless communications and radar systems are identified as part of future scope of this work.
\new{In the future, we envision extending DS3 to other domains with appropriate case studies and validation.}


%% file: files/biographies.tex
\begin{IEEEbiography}
[{\includegraphics[width=1in,height=1.25in,clip,keepaspectratio]{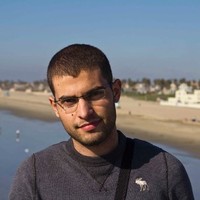}}]{Samet E. Arda} received his M.S. and Ph.D. degrees in School of Electrical, Computer, and Energy Engineering from Arizona State University (ASU). He is currently an Assistant Research Scientist at ASU.
\end{IEEEbiography}
\vskip -2.5\baselineskip plus -1fil
\begin{IEEEbiography}
[{\includegraphics[width=1in,height=1.25in,clip,keepaspectratio]{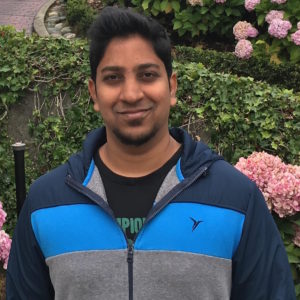}}]{Anish NK} received his B.Tech degree in Electrical and Electronics Engineering from National Institute of Technology, Tiruchirappalli, India and Masters degree from Birla Institute of Technology, Pilani, India. Anish worked as a Physical Design Engineer at Qualcomm and as a Research Scientist at Intel Labs in India. He is currently pursuing Ph.D in Electrical Engineering at Arizona State University.
\end{IEEEbiography}
\vskip -2.5\baselineskip plus -1fil
\begin{IEEEbiography}
[{\includegraphics[width=1in,height=1.25in,clip,keepaspectratio]{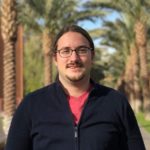}}]{A. Alper Goksoy} received his B.S. degree in Electrical and Electronics Engineering from Bogazici University, Istanbul, Turkey.  He is currently pursuing his Ph.D. in Electrical Engineering at Arizona State University.
\end{IEEEbiography}
\vskip -2.5\baselineskip plus -1fil
\begin{IEEEbiography}
[{\includegraphics[width=1in,height=1.25in,clip,keepaspectratio]{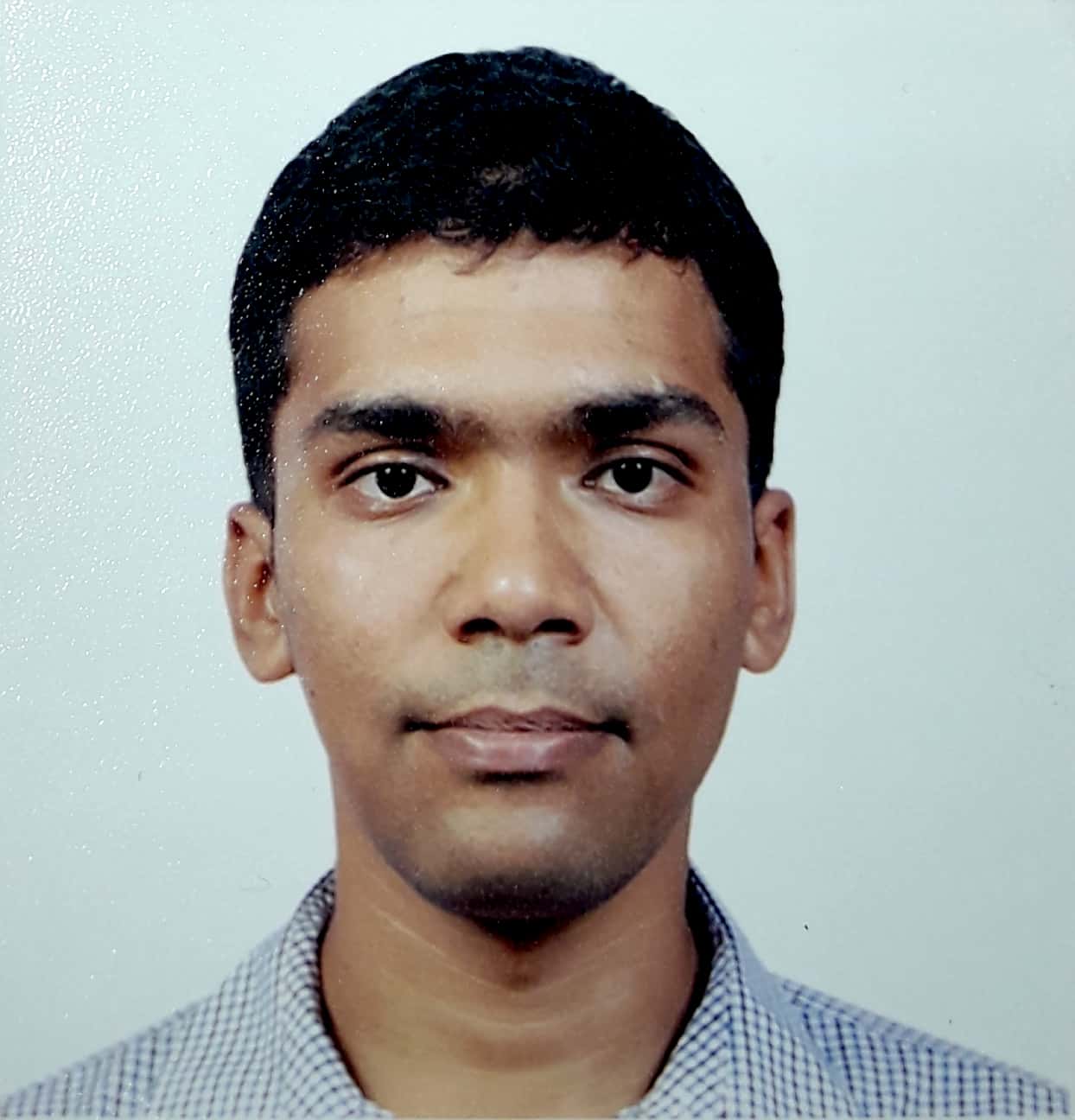}}]{Nirmal Kumbhare} Nirmal Kumbhare received his Ph.D. in computer engineering from the University of Arizona. His research interests involve reconfigurable and heterogeneous systems, high performance computing, and power-aware resource management. He has worked with Intel India prior to joining Ph.D.
\end{IEEEbiography}
\vskip -2.5\baselineskip plus -1fil
\begin{IEEEbiography}
[{\includegraphics[width=1in,height=1.25in,clip,keepaspectratio]{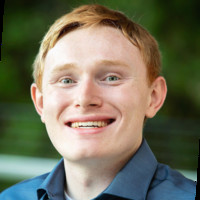}}]{Joshua Mack} Joshua Mack is a doctoral student in the University of Arizona Electrical and Computer Engineering program. His research interests include the intersection of high performance computing and reconfigurable systems; emerging architectures; and intelligent and/or autonomous workload partitioning across heterogeneous systems.
\end{IEEEbiography}
\vskip -2.5\baselineskip plus -1fil
\begin{IEEEbiography}
[{\includegraphics[width=1in,height=1.25in,clip,keepaspectratio]{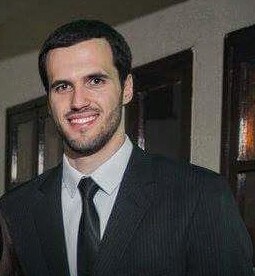}}]{Anderson L. Sartor} received his B.Sc. and Ph.D. in Computer Engineering from Universidade Federal do Rio Grande do Sul (UFRGS), Brazil, in 2013 and 2018, respectively. He is currently a Postdoctoral Researcher at Carnegie Mellon University (CMU). His primary research interests include embedded systems design, machine learning for energy optimization, SoC modeling, and adaptive processors.
\end{IEEEbiography}
\vskip -2.5\baselineskip plus -1fil
\begin{IEEEbiography}
[{\includegraphics[width=1in,height=1.25in,clip,keepaspectratio]{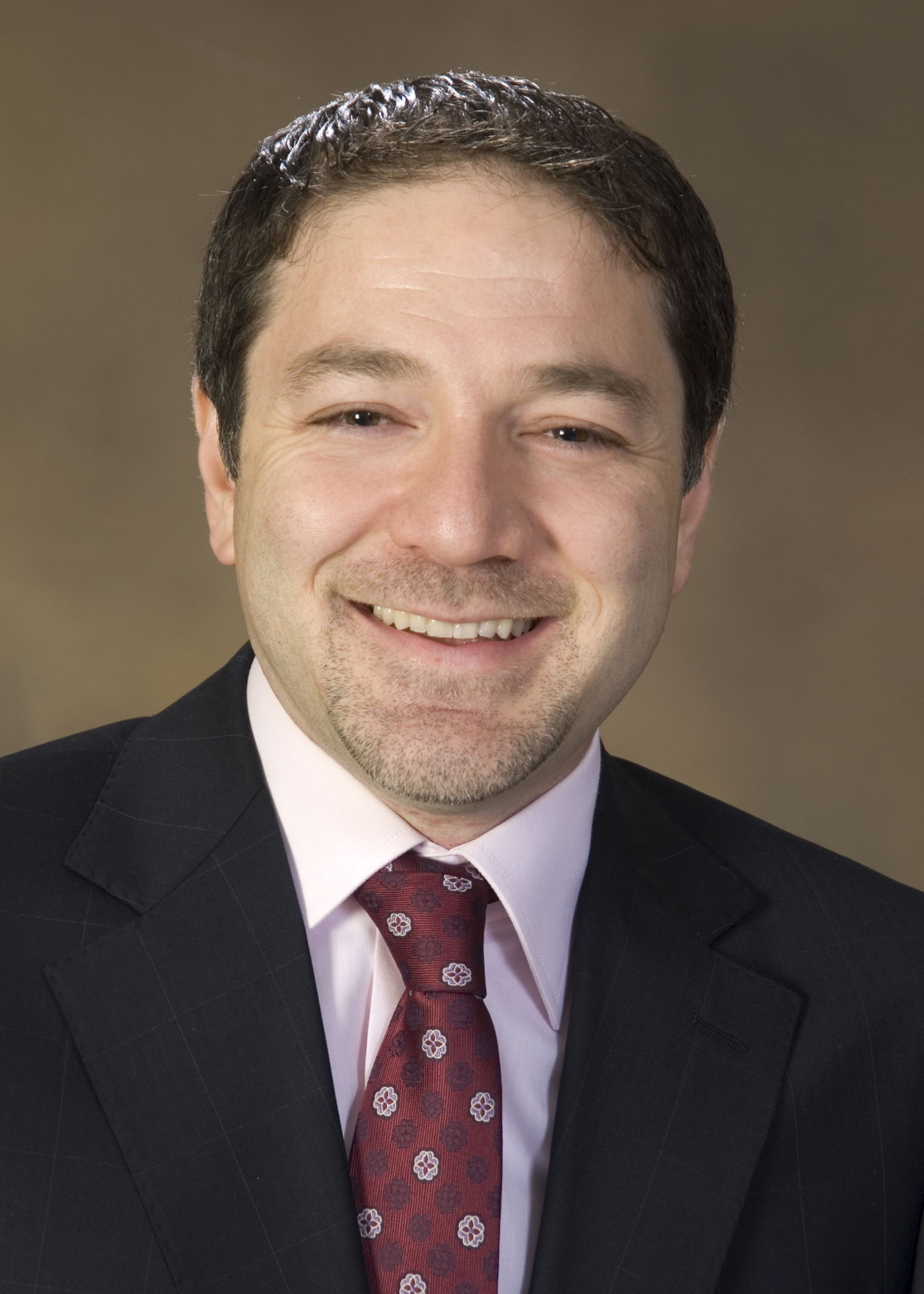}}]{Ali Akoglu} 
received his Ph.D. degree in Computer Science from the Arizona State University in 2005. He is an Associate Professor in the Department of Electrical and Computer Engineering and the BIO5 Institute at the University of Arizona. He is the site-director of the National Science Foundation Industry-University Cooperative Research Center on Cloud and Autonomic Computing. His research program focuses on high performance computing and non-traditional computing architectures. 
\end{IEEEbiography}
\vskip -3.0\baselineskip plus -1fil
\begin{IEEEbiography}
[{\includegraphics[width=1in,height=1.25in,clip,keepaspectratio]{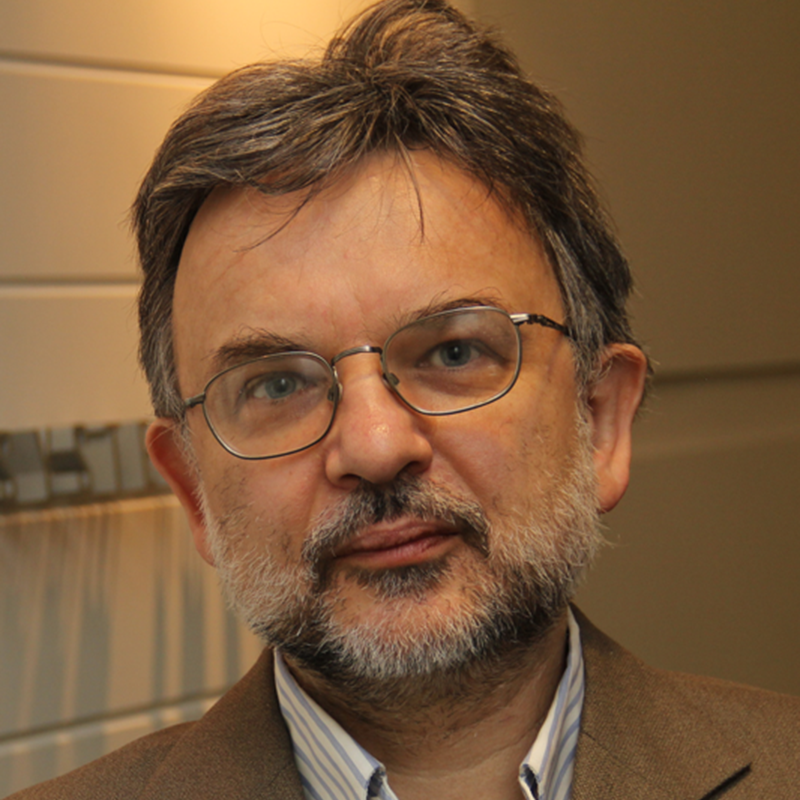}}]{Radu Marculescu} is the Kavčić-Moura Professor of Electrical and Computer Engineering at Carnegie Mellon University. He received his Ph.D. in Electrical Engineering from the University of Southern California in 1998. Radu's current research focuses on developing methods and tools for modeling and optimization of embedded systems, cyber-physical systems, social networks, and biological systems.
\end{IEEEbiography}
\vskip -2.5\baselineskip plus -1fil
\begin{IEEEbiography}
[{\includegraphics[width=1in,height=1.25in,clip,keepaspectratio]{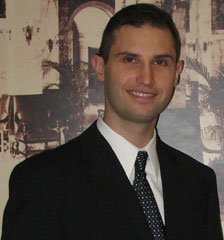}}]{Umit Y. Ogras} received his Ph.D. degree in Electrical and Computer Engineering from Carnegie Mellon University, Pittsburgh, PA, in 2007. From 2008 to 2013, he worked as a research scientist at the Strategic CAD Laboratories, Intel Corporation. He is an Associate Professor at the School of Electrical, Computer and Energy Engineering, and the Associate Director of WISCA Center.
\end{IEEEbiography}